\documentclass[11pt]{amsart}

\usepackage[margin=1.1in]{geometry}
\usepackage{amssymb}
\usepackage{mathtools}
\usepackage{booktabs}
\usepackage{tikz}
\usetikzlibrary{calc}
\usepackage{microtype}
\usepackage{etoolbox}
\PassOptionsToPackage{hyphens}{url}
\usepackage[colorlinks=true,linkcolor=blue!55!black,citecolor=green!40!black,urlcolor=blue!55!black]{hyperref}

\newtheorem{theorem}{Theorem}[section]
\newtheorem{proposition}[theorem]{Proposition}
\newtheorem{lemma}[theorem]{Lemma}
\newtheorem{corollary}[theorem]{Corollary}
\newtheorem{mainthm}{Theorem}

\theoremstyle{definition}
\newtheorem{definition}[theorem]{Definition}
\newtheorem{example}[theorem]{Example}
\newtheorem{remark}[theorem]{Remark}
\newtheorem{question}[theorem]{Question}

\numberwithin{equation}{section}

\newcommand{\C}{\mathbb{C}}
\newcommand{\Z}{\mathbb{Z}}
\newcommand{\Q}{\mathbb{Q}}

\newcommand{\cS}{\mathcal{S}}
\newcommand{\cF}{\mathcal{F}}
\DeclareMathOperator{\tr}{tr}

\DeclareMathOperator{\id}{id}

\newcommand{\fc}[4]{\left[\begin{smallmatrix}#1&#2\\#3&#4\end{smallmatrix}\right]}
\newcommand{\bz}{\mathtt{0}}
\newcommand{\bL}{\mathtt{L}}
\newcommand{\bR}{\mathtt{R}}

\makeatletter
\let\@affiliations\@empty
\newcommand{\affiliation}[1]{\g@addto@macro\@affiliations{\par#1}}
\patchcmd{\@setauthors}{\endtrivlist}
  {\par\smallskip{\let\\\par\@affiliations\par}\endtrivlist}
  {}{\errmessage{Could not add the affiliations to the title block}}
\g@addto@macro\addresses{\author{}}
\makeatother

\begin{document}

\title[A Yang--Baxter obstruction for hard squares]{A local Yang--Baxter obstruction\\ for the hard-square model}

\author[Gabriel Ortega]{Gabriel Ortega\textsuperscript{1,2}}
\email{gabriel.ortega@ciae.uchile.cl}
\author[Eliecer Pe\~na-Ancavil]{Eliecer Pe\~na-Ancavil\textsuperscript{2,3}}
\affiliation{\textsuperscript{1} Centro de Investigaci\'on Avanzada en Educaci\'on (CIAE)\\
  Instituto de Estudios Avanzados en Educaci\'on (IE)\\
  Universidad de Chile}
\affiliation{\textsuperscript{2} Diagnostika Labs, Santiago, Chile}
\affiliation{\textsuperscript{3} Universidad Bernardo O'Higgins, Santiago, Chile}


\begin{abstract}
Many exact solutions of two-dimensional lattice models rest on an analytic family of weights satisfying the Yang--Baxter equation. The hard-hexagon model was solved in this way, while the neighbouring hard-square model has no known closed form. We determine the height alphabets that encode hard squares without hidden heights, and there are exactly two connected ones, the occupation variables themselves and the four heights of Andrews, Baxter and Forrester. Hard hexagons is integrable on both, but no analytic Yang--Baxter family of face weights passes through hard squares at nonzero activity on either. The difference lies in the exclusion rule, which for hard squares turns a few Yang--Baxter equations into polynomial identities that force auxiliary weights to vanish. A closed form for hard squares is not excluded, but a solution by this route would have to leave these encodings or go beyond a single analytic family through the model. Encodings with hidden heights remain open.
\end{abstract}

\maketitle

\setcounter{tocdepth}{1}
\tableofcontents

\section{Introduction}\label{sec:intro}

\subsection{Hard hexagons and hard squares}

The \emph{hard-square model} is the lattice gas on $\Z^2$ with nearest-neighbour exclusion. A configuration is a map $x\colon\Z^2\to\{0,1\}$ with $x_vx_{v+e_1}=x_vx_{v+e_2}=0$ for all $v$, and every occupied site carries an activity $z$. Its free energy can be computed to high precision \cite{BaxterEntingTsang80,Baxter99,Chan12}, but no closed form is known.

The neighbouring \emph{hard-hexagon model}, in which the exclusion also acts along one diagonal of each plaquette, was solved exactly by Baxter \cite{Baxter80,BaxterBook}. The solution rests on the model being a specialization of a family of face weights, depending analytically on a spectral parameter, that satisfies the face Yang--Baxter equation, so that row transfer matrices with different spectral parameters commute. Baxter solved it directly in the occupation variables, as a special case of hard squares with diagonal interactions \cite{Baxter80}. In the formulation of Andrews, Baxter and Forrester \cite{ABF84}, an integrable family through hard hexagons lives on the four heights $1,2,3,4$ of a restricted solid-on-solid model; the occupation of a site is read off from its height, occupied sites being those of height $1$ or $4$. Hard hexagons is thus the shadow, under the occupation map, of an integrable height model in two ways, on $A_4$ and on the binary alphabet of occupation variables of Section~\ref{sec:main-result}, so the heights of $A_4$ are not needed to solve it.

Hard squares is widely believed not to be integrable \cite{Pearce16}. The belief rests on the absence of a solution despite long series and transfer-matrix studies \cite{GauntFisher65,RunnelsCombs66,BaxterEntingTsang80,Chan12}, on the comparison of partition-function zeros and transfer-matrix eigenvalues with those of hard hexagons \cite{AJJMM14}, and on the observation that Baxter's integrable family of hard squares with diagonal interactions \cite{Baxter80,Baxter81,Huse82,BaxterPearce83} meets the pure hard-square model only at zero activity, as noted in \cite[after Eq.~(24)]{Baxter80} and \cite[Appendix~B]{TanakaTamuraKatsura12} (Remark~\ref{rem:baxter-surface}). None of this excludes that hard squares, like hard hexagons, is the shadow of some integrable height model. This paper addresses exactly that question.

\subsection{The main result}\label{sec:main-result}

A \emph{height alphabet} is a finite set of heights with a nearest-neighbour adjacency rule and an occupation map to $\{0,1\}$. It \emph{encodes hard squares without hidden heights} if, on every even torus, every hard-square configuration is the image of the same bounded number of height configurations (Definition~\ref{def:encoding}). The binary alphabet $\{0,1\}$ itself is such an encoding, and so is the hard-hexagon alphabet $A_4$ above, whose heights record the occupation of a site together with the parity of its sublattice, so that every hard-square configuration has two lifts. Our first result says that these are the only ones.

\begin{mainthm}[Classification; Proposition~\ref{prop:classification}]\label{thmA}
A connected height alphabet that encodes hard squares without hidden heights is isomorphic either to the binary alphabet or to the alphabet $A_4$ of Andrews, Baxter and Forrester with occupied heights $\{1,4\}$.
\end{mainthm}

The second case is the alphabet behind the exact solution of hard hexagons. The two-to-one lift of hard-square configurations to the path $1-2-3-4$ is known \cite{FendleySenguptaSachdev04,ChandgotiaMarcus18}; what Theorem~\ref{thmA} adds is that there is no other case, without assuming that the occupation map is a covering.

A face model on such an alphabet \emph{presents} hard squares if every height configuration receives exactly the hard-square weight of its image. We determine all such presentations (Proposition~\ref{prop:presentations}); on $A_4$ they are parity-staggered versions of the binary hard-square face. On these two alphabets, requiring only that the height weights reproduce the hard-square weight after summing over the fibres of the occupation map gives nothing new (Proposition~\ref{prop:fibre-sums}). The central result is that none of these presentations can be deformed into a Yang--Baxter family.

\begin{mainthm}[No Yang--Baxter family through hard squares; Theorem~\ref{thm:main}]\label{thmB}
Let the height alphabet be the binary alphabet or $A_4$. There is no family $(W_{uv})$ of face weights on it, depending analytically on spectral parameters $u,v$ in a connected domain, such that every triple $(W_{pq},W_{pr},W_{qr})$ satisfies the face Yang--Baxter equation, even up to a scalar factor, and such that some $W_{p_0r_0}$ is a presentation of hard squares of nonzero activity.
\end{mainthm}

Theorem~\ref{thmB} is sharp on both alphabets, since with the same occupation map each of them carries a Yang--Baxter family through hard hexagons. On the binary alphabet this is Baxter's family, whose weights are holomorphic functions of the spectral parameter $w\neq0$ and give hard hexagons at $w=x^2$ \cite[Eq.~(28a)]{Baxter80}, and on $A_4$ it is the family of \cite{ABF84}. At the critical activity, Example~\ref{ex:rsos-calibration} gives such families explicitly on both alphabets. What separates the two models is the hard-core rule itself. What $A_4$ adds is the staggered presentations, which have no binary counterpart. Remark~\ref{rem:why-class} explains why encodings without hidden heights form a natural class. On them, a commuting family of transfer matrices through a presentation would contain, up to a scalar and a diagonal similarity, the hard-square transfer matrix itself, or on $A_4$ a matrix whose square is two copies of its square.

\subsection{The mechanism}

The proof is algebraic and local. When the middle face of the Yang--Baxter equation is a hard-square face, the exclusion rule kills most terms of certain boundary equations. A short combination of seven of them (eight on $A_4$) yields an explicit polynomial certificate, which forces the product of three auxiliary weights to vanish:
\[
 r_5\,p_3\,r_2=0\quad\text{(binary alphabet)},\qquad
 r^{(0)}_5\,p^{(0)}_2\,r^{(0)}_3=0\quad\text{(alphabet $A_4$)}
\]
(Theorems~\ref{thm:binary-certificate} and~\ref{thm:a4-certificate}). In a spectral family, each auxiliary face coincides with the hard-square face at one endpoint of the spectral parameter, where none of these weights vanishes; analyticity then gives a contradiction. The certificates are division-free up to the corner factors of the hard-square face, and every step can be checked by hand.

\subsection{Sharpness and extensions}

Theorem~\ref{thmB} detects the pure hard-core rule, and it is sharp with respect to the family hypothesis.
\begin{itemize}
\item Adding diagonal interactions changes the model and restores integrability, since hard hexagons on $A_4$ and Baxter's diagonal-interaction surface are specializations of Yang--Baxter families.
\item The obstruction needs the auxiliary faces to lie in a family with the middle face, or one of them to have full support. A single Yang--Baxter equation \eqref{eq:FYB} on the binary alphabet with a hard-square middle face has solutions, in which at least two auxiliary weights vanish, and those with exactly two zeros form two explicit toric families (Theorem~\ref{thm:near-full}). When one auxiliary face has full support, these trivial solutions are the only ones, whether or not the faces belong to a family (Proposition~\ref{prop:rigidity}).
\end{itemize}
The same mechanism goes beyond height alphabets (Section~\ref{sec:extensions}). Blocking two sites into one gives a three-letter alphabet with $41$ admissible faces; there a \emph{single} boundary equation already excludes every analytic Yang--Baxter family of weights on these faces through hard squares, with no genericity assumption (Corollary~\ref{cor:blocked-analytic}). This concerns \eqref{eq:FYB} with all three faces on the admissible blocked faces; the transfer-matrix structure of the blocked model, in which the intertwining face satisfies the vertical rule on its four edges, is not covered. On the binary alphabet with $1\sim1$, where the hard-core rule is imposed by a hard-square face that vanishes on every face with two adjacent occupied sites, twelve boundary equations again exclude every analytic family (Proposition~\ref{prop:full-binary}).

\subsection{Scope}\label{sec:scope}

Commuting transfer matrices obtained from a local Yang--Baxter family are the classical route to a closed formula for the free energy of a two-dimensional lattice model. Theorems~\ref{thmA} and~\ref{thmB} and Corollary~\ref{cor:reduced} show that, for hard squares, this mechanism is unavailable on every height alphabet that encodes the model without hidden heights, including $A_4$, in the form of Definition~\ref{def:analytic-family}, that is, as a single analytic family of face weights that passes through a presentation of hard squares at a point of its connected parameter domain. They do not exclude hard squares as a limit or a boundary value of such a family, a transfer matrix that commutes with a family without belonging to it, such as the hard-square transfer matrix intertwined with the row transfer matrices of a family by faces outside the family (on the binary alphabet, Proposition~\ref{prop:rigidity} excludes this for rows of full support other than pure hard-square faces), or constructions with different families on different lines, such as lines of different fusion levels or intertwiners with another model.

Nor do they cover, in general, alphabets in which occupied heights may be adjacent and the hard-core rule is imposed only by weights that vanish at the special point; edge-interaction formulations with a star--triangle relation; vertex--face correspondences with auxiliary spaces; functional relations that do not come from a family of face weights; or reformulations that are not encodings in the sense of Definition~\ref{def:encoding}. Two reformulations of this kind are natural. Summing over one sublattice gives a face model on the other, a square lattice rotated by $45^\circ$, on the binary alphabet with all sixteen faces admissible, with weight $1+z$ on the empty face and $1$ on the other faces, and the activity $z$ on its sites \cite[Section~3]{Baxter11}. In a checkerboard colouring of the plaquettes of $\Z^2$, every site is a corner of two black plaquettes and every edge lies in exactly one, so hard squares becomes a vertex model on the lattice of black plaquettes. The sites of $\Z^2$ are its edges, the vertex configurations are the seven independent sets of the four corners of a plaquette, and each occupied corner has weight $z^{1/2}$; its integrability would go through the vertex Yang--Baxter equation. Section~\ref{sec:extensions} treats two cases outside Definition~\ref{def:encoding}, namely blocks of two sites, with all three faces of the equation on the admissible blocked faces, and the binary alphabet with $1\sim1$ when the special face vanishes on every face with two adjacent occupied sites (Proposition~\ref{prop:full-binary}).

The results do not show that the free energy of hard squares has no closed form, since closed formulas can arise from other mechanisms, as for the dimer model \cite{Kasteleyn61,TemperleyFisher61}. The open case is that of encodings with hidden heights, where several height configurations of unbounded number lie over a single hard-square configuration; we return to it in Section~\ref{sec:conclusion}.

\subsection{Related work}

Face Yang--Baxter equations, the star--triangle relation and commuting transfer matrices are classical \cite{BaxterBook,Baxter72}; restricted solid-on-solid models \cite{ABF84}, fusion \cite{DateEtAl87,ZhouPearce94} and vertex--face correspondences \cite{Pasquier88} produce the standard integrable face models, and Baxterization \cite{Jones90} produces spectral families from braid-group representations. Instead of constructing solutions, we certify that none passes through a prescribed point. Hard squares and hard hexagons are compared through partition-function zeros and transfer-matrix eigenvalues in \cite{AJJMM14}; long series for hard squares are computed in \cite{Chan12}, and the diagonal-interaction model is studied with tensor networks in \cite{NyckeesMila22}. For quantum spin chains, nonintegrability has been proved by excluding local conserved quantities \cite{Shiraishi19}, also for the PXP chain \cite{ParkLee25}, whose kinetic constraint is the one-dimensional hard-core rule, and Yang--Baxter integrability has been characterized by the Reshetikhin condition \cite{SanataniShiraishiIshii26}; these results concern one-dimensional quantum Hamiltonians, not face weights of a two-dimensional model.

\subsection{Organization}

Section~\ref{sec:prelim} introduces faces, the face Yang--Baxter equation and height alphabets, and proves Theorem~\ref{thmA} together with the classification of presentations. Section~\ref{sec:support} proves the certificates and Theorem~\ref{thmB}, and discusses sharpness. Section~\ref{sec:extensions} treats blocks of two sites and the binary alphabet with $1\sim1$. Section~\ref{sec:conclusion} discusses hidden heights and open questions. Appendix~\ref{app:computations} lists the finite computations, all reproduced by an ancillary script.

\section{Height alphabets and presentations of hard squares}\label{sec:prelim}

Throughout, vector spaces are finite dimensional over $\C$, and $\tr$ denotes the trace of an endomorphism.

\subsection{Faces and the face Yang--Baxter equation}

A \emph{height alphabet} is a finite set $\cS$ with a symmetric relation $\sim$ (loops allowed), the \emph{adjacency rule}. A \emph{face} is an array $\fc abcd$ of heights, with top row $a,b$, bottom row $c,d$, left column $(a,c)$ and right column $(b,d)$. It is \emph{admissible} if $a\sim b$, $c\sim d$, $a\sim c$ and $b\sim d$; the pairs $\{a,d\}$ and $\{b,c\}$ are its diagonals, on which no condition is imposed. A \emph{face weight} is a function on admissible faces, extended by zero to all other faces.

\begin{definition}[Face Yang--Baxter equation]\label{def:fyb}
Face weights $P,Q,R$ on $\cS$ satisfy the \emph{face Yang--Baxter equation} if, for all boundary heights $(a,b,c,d,e,f)\in\cS^6$,
\begin{equation}\label{eq:FYB}
 \sum_{g\in\cS}P\fc fgab\,Q\fc edfg\,R\fc dcgb
 \;=\;
 \sum_{g\in\cS}R\fc egfa\,Q\fc gcab\,P\fc edgc .
\end{equation}
They satisfy it \emph{up to a scalar factor} $\rho\in\C^\times$ if the right side is multiplied by $\rho$:
\begin{equation}\label{eq:FYB-rho}
 \sum_{g}P\fc fgab\,Q\fc edfg\,R\fc dcgb
 \;=\;\rho\sum_{g}R\fc egfa\,Q\fc gcab\,P\fc edgc .
\end{equation}
When the weights take values in an associative algebra, the products are read in the displayed order.
\end{definition}

\begin{figure}[ht]
\centering
\begin{tikzpicture}[scale=1.15,every node/.style={font=\small}]
 \begin{scope}
  \coordinate (a) at (0,0); \coordinate (b) at (2,0); \coordinate (c) at (3,1.2);
  \coordinate (d) at (2,2.4); \coordinate (e) at (0,2.4); \coordinate (f) at (-1,1.2);
  \coordinate (g) at (1,1.2);
  \draw[thick] (a)--(b)--(c)--(d)--(e)--(f)--cycle;
  \draw (f)--(g)--(b); \draw (g)--(d);
  \node[below left] at (a) {$a$}; \node[below right] at (b) {$b$}; \node[right] at (c) {$c$};
  \node[above right] at (d) {$d$}; \node[above left] at (e) {$e$}; \node[left] at (f) {$f$};
  \node[below] at ($(g)+(0,-0.05)$) {$g$};
  \node at (0.5,0.6) {$P$}; \node at (0.5,1.8) {$Q$}; \node at (2.0,1.2) {$R$};
 \end{scope}
 \node at (4.0,1.2) {$=$};
 \begin{scope}[xshift=6cm]
  \coordinate (a) at (0,0); \coordinate (b) at (2,0); \coordinate (c) at (3,1.2);
  \coordinate (d) at (2,2.4); \coordinate (e) at (0,2.4); \coordinate (f) at (-1,1.2);
  \coordinate (g) at (1,1.2);
  \draw[thick] (a)--(b)--(c)--(d)--(e)--(f)--cycle;
  \draw (a)--(g)--(c); \draw (g)--(e);
  \node[below left] at (a) {$a$}; \node[below right] at (b) {$b$}; \node[right] at (c) {$c$};
  \node[above right] at (d) {$d$}; \node[above left] at (e) {$e$}; \node[left] at (f) {$f$};
  \node[above] at ($(g)+(0,0.05)$) {$g$};
  \node at (0.0,1.2) {$R$}; \node at (1.5,0.6) {$Q$}; \node at (1.5,1.8) {$P$};
 \end{scope}
\end{tikzpicture}
\caption{The two tilings of the hexagon underlying \eqref{eq:FYB}. On the left, $P$ occupies the rhombus $(a,b,g,f)$, $Q$ the rhombus $(f,g,d,e)$ and $R$ the rhombus $(g,b,c,d)$; on the right, $R$ occupies $(f,a,g,e)$, $Q$ occupies $(a,b,c,g)$ and $P$ occupies $(g,c,d,e)$. In an analytic Yang--Baxter family (Definition~\ref{def:analytic-family}), $P$, $Q$ and $R$ are $W_{pq}$, $W_{pr}$ and $W_{qr}$.}
\label{fig:hexagon}
\end{figure}
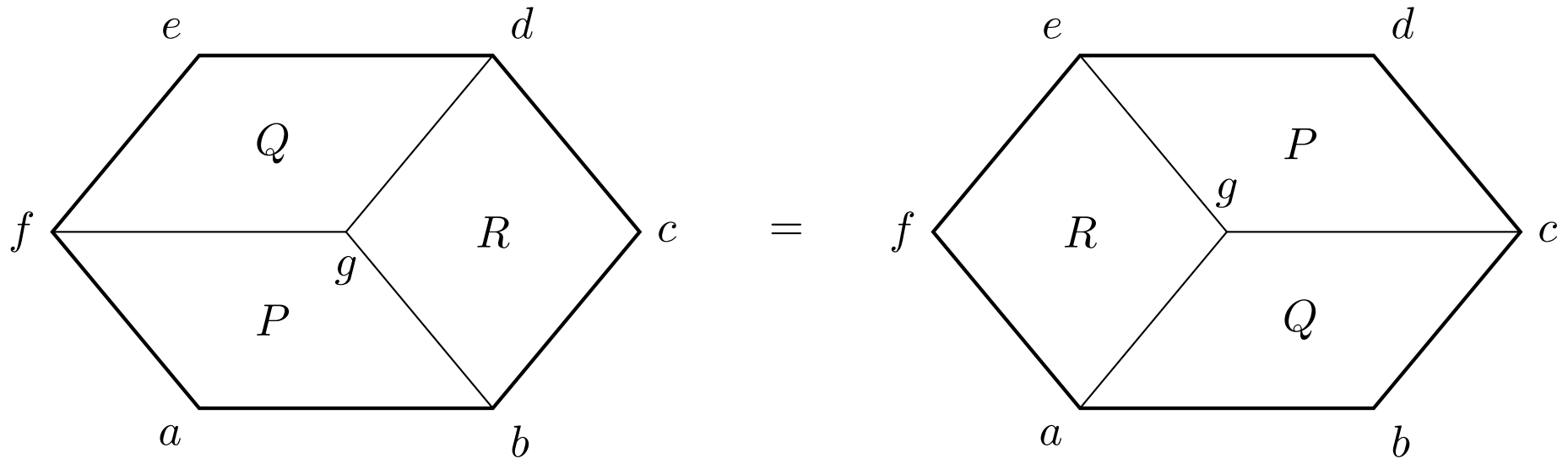

Equation \eqref{eq:FYB} is the interaction-round-a-face form of the Yang--Baxter relation \cite{BaxterBook}, also called its star--triangle form; it is distinct from the star--triangle relation of models with edge interactions. The six boundary heights sit at the vertices of a hexagon, $g$ is an interior height, and the two sides correspond to the two tilings of the hexagon by three rhombi (Figure~\ref{fig:hexagon}). In both tilings the rhombus of type $P$ has edges parallel to $ab$ and $af$, that of type $Q$ to $ab$ and $bc$, and that of type $R$ to $bc$ and $af$. The convention is the standard one. It is, symbol by symbol, Eq.~(A.5) of \cite{WarnaarNienhuis93}, and it is satisfied by the restricted solid-on-solid models, as recalled in Example~\ref{ex:rsos-calibration}.

\begin{definition}[Yang--Baxter families]\label{def:analytic-family}
Let $\mathcal D\subset\C$ be a connected open set. An \emph{analytic Yang--Baxter family} on $(\cS,\sim)$ is a family $(W_{uv})_{u,v\in\mathcal D}$ of face weights on $\cS$, holomorphic in $(u,v)$, such that \eqref{eq:FYB-rho} holds for $(P,Q,R)=(W_{pq},W_{pr},W_{qr})$ for all $p,q,r\in\mathcal D$, with a nowhere-vanishing scalar factor $\rho=\rho(p,q,r)$. It \emph{passes through} a face weight $Q_0$ if $W_{p_0r_0}=Q_0$ for some $p_0,r_0\in\mathcal D$.
\end{definition}

No difference property, unitarity or regularity is assumed. Families of difference form, in which $W_{uv}$ depends only on $v-u$, are a special case. For example, the weights $W_t$ of Example~\ref{ex:rsos-calibration} give the family $W_{uv}=W_{v-u}$, for which $(W_{pq},W_{pr},W_{qr})=(W_u,W_{u+v},W_v)$ with $u=q-p$ and $v=r-q$. The only feature of Definition~\ref{def:analytic-family} that we use is that, with $p=p_0$ and $r=r_0$ fixed, the auxiliary faces $P=W_{p_0q}$ and $R=W_{qr_0}$ depend holomorphically on $q$ and reduce to the middle face at $q=r_0$ and $q=p_0$ respectively. Since the analytic part of the proofs restricts the family to $q\mapsto(p_0,q,r_0)$ and uses only the identity theorem, it suffices that \eqref{eq:FYB-rho} hold for these triples with $q$ in a nonempty open subset of $\mathcal D$, and the proofs apply verbatim to meromorphic families, such as elliptic ones, and to rapidities on a connected Riemann surface. A family that satisfies \eqref{eq:FYB} only up to factors depending on the boundary heights is covered when it is a gauge transform of a family as in Definition~\ref{def:analytic-family}, since gauge transformations cancel on tori and preserve presentations. Commuting transfer matrices follow from \eqref{eq:FYB-rho} by the train argument under the usual nondegeneracy of the intertwining face; the proofs below use only the local equations. In that argument $P$ and $Q$ are faces of two consecutive rows, the top row of $P$ being the bottom row of $Q$, and $R$ is the intertwining face, each of whose edges is a column of $P$ or of $Q$ on one of the two sides of \eqref{eq:FYB}. In Theorem~\ref{thm:main} the hard-square face is thus the row face $W_{p_0r_0}$, and the intertwiner $W_{qr_0}$ belongs to the family; intertwiners that belong to no family are treated, on the binary alphabet, in Proposition~\ref{prop:rigidity}.

\subsection{Height alphabets that encode hard squares}

A \emph{height alphabet with occupation} is a height alphabet $(\cS,\sim)$ together with a map $n\colon\cS\to\{0,1\}$; write $\cS=U\sqcup O$ with $U=n^{-1}(0)$ (empty heights) and $O=n^{-1}(1)$ (occupied heights). For $L\ge2$ let $T_L=(\Z/L\Z)^2$ be the torus. A \emph{height configuration} on $T_L$ is a map $h\colon T_L\to\cS$ with $h(v)\sim h(w)$ for all nearest neighbours $v,w$; its \emph{image} is $n\circ h\in\{0,1\}^{T_L}$.

\begin{definition}[Encodings without hidden heights]\label{def:encoding}
A height alphabet with occupation \emph{encodes hard squares without hidden heights} if
\begin{enumerate}
\item no occupied height is adjacent to an occupied height, itself included, so that every image is a hard-square configuration;
\item for every even $L$ there is $\kappa_L\ge1$ such that every hard-square configuration on $T_L$ is the image of exactly $\kappa_L$ height configurations, and $\sup_L\kappa_L<\infty$.
\end{enumerate}
It is \emph{reduced} if every height occurs in some height configuration on some even torus, and \emph{connected} if the graph $(\cS,\sim)$ is connected.
\end{definition}

Condition (2) says that the height model carries no degrees of freedom beyond the hard-square configuration and a bounded global choice. Hidden refers here to heights. Auxiliary spaces, gauge factors or bond variables that disappear under a trace are a different matter, not treated by Definition~\ref{def:encoding}. Only even tori enter because $A_4$ is bipartite, so it has no height configurations on odd tori and asking (2) for every $L$ would exclude it. The occupation of a site is read from its own height. Encodings that read it from several heights, or that group several sites into one height, fall outside Definition~\ref{def:encoding}; Section~\ref{sec:extensions} treats blocks of two sites. So do reformulations on other lattices, such as those described in Section~\ref{sec:scope}. A connected alphabet with at least two heights is reduced, since a height $h$ with a neighbour $h'$ occurs in the configuration equal to $h$ on one sublattice and to $h'$ on the other; an alphabet with a single height does not satisfy (2).

\begin{figure}[ht]
\centering
\begin{tikzpicture}[scale=0.7,every node/.style={font=\small}]
 \draw[gray!60] (0,0) grid (3,3);
 \foreach \x/\y in {0/0,2/0,3/1,1/2,0/3,3/3} \fill (\x,\y) circle (0.17);
 \foreach \x/\y in {1/0,3/0,0/1,1/1,2/1,0/2,2/2,3/2,1/3,2/3} \draw[fill=white] (\x,\y) circle (0.09);
 \draw[->] (3.7,1.5) -- (4.3,1.5);
 \begin{scope}[shift={(5,0)}]
  \draw[gray!60] (0,0) grid (3,3);
  \foreach \x/\y/\h in {0/0/1,1/0/2,2/0/1,3/0/2,0/1/2,1/1/3,2/1/2,3/1/1,0/2/3,1/2/4,2/2/3,3/2/2,0/3/4,1/3/3,2/3/2,3/3/1}
   \node[fill=white,inner sep=1pt] at (\x,\y) {$\h$};
 \end{scope}
\end{tikzpicture}
\caption{A hard-square configuration on a $4\times4$ patch (left, occupied sites in black) and one of its two lifts to $A_4$ (right), with odd heights on the sublattice of the bottom-left site. An occupied site carries $1$ or $4$, and an empty site $3$ or $2$, according to its sublattice. The patch contains isolated particles and occupied diagonal pairs of both kinds. The other lift is obtained by $k\mapsto5-k$.}
\label{fig:lift}
\end{figure}
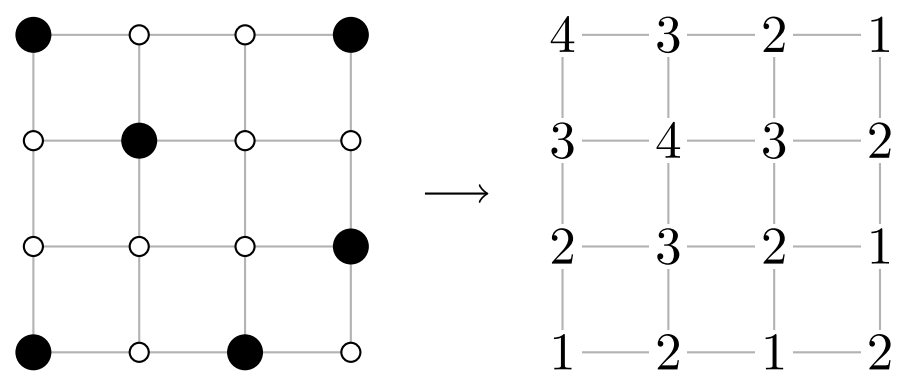

\begin{example}\label{ex:encodings}
\begin{enumerate}
\item The \emph{binary alphabet} has $\cS=\{0,1\}$, $0\sim0$, $0\sim1$ and $n=\id$. Here $\kappa_L=1$.
\item The alphabet $A_4$ has $\cS=\{1,2,3,4\}$, with $h\sim h'$ if and only if $|h-h'|=1$, and $n(1)=n(4)=1$, $n(2)=n(3)=0$. Adjacent heights have opposite parity, so on an even torus the heights of one sublattice are odd and those of the other are even. Given a hard-square configuration $x$ and a choice of the odd sublattice, the heights are forced (Figure~\ref{fig:lift}). An occupied (empty) site gets height $1$ ($3$) on the odd sublattice and $4$ ($2$) on the even one; the adjacency rule then holds exactly because no two neighbours are occupied. Hence $\kappa_L=2$, and the two lifts are exchanged by the reflection $\sigma(k)=5-k$ of the heights. This is the height alphabet of the restricted solid-on-solid model of Andrews, Baxter and Forrester with $r=5$, on which the hard-hexagon model is solvable \cite{ABF84,BaxterBook}; the same height correspondence for hard squares is spelled out in \cite[Appendix]{FendleySenguptaSachdev04}.
\end{enumerate}
\end{example}

\begin{proposition}[Classification]\label{prop:classification}
A connected reduced height alphabet with occupation that encodes hard squares without hidden heights is isomorphic to the binary alphabet or to $A_4$. More generally, a reduced encoding without hidden heights is a disjoint union of copies of these two.
\end{proposition}

\begin{proof}
Fix an even $L\ge4$ with $2^{L^2/2}>\sup_M\kappa_M$ and let $\Lambda_0,\Lambda_1$ be the two sublattices of $T_L$. Write $\kappa=\kappa_L$.

\emph{Step 1: occupied and empty heights are matched.} The configuration $x_\Lambda=\mathbf 1_{\Lambda_0}$ is a hard-square configuration. If an occupied height $o$ had two distinct empty neighbours $u\ne u'$, then $h\equiv o$ on $\Lambda_0$, with $h$ arbitrary in $\{u,u'\}$ on $\Lambda_1$, would give $2^{L^2/2}$ preimages of $x_\Lambda$. If an empty height $u$ had two distinct occupied neighbours $o\neq o'$, then $h\equiv u$ on $\Lambda_1$, with $h$ arbitrary in $\{o,o'\}$ on $\Lambda_0$, would again give $2^{L^2/2}$ preimages of $x_\Lambda$. Since $\kappa<2^{L^2/2}$, the adjacency between $O$ and $U$ is a partial matching. Every $o\in O$ occurs at some site, whose four neighbours carry empty heights adjacent to $o$; hence every $o$ is matched, to a height we call $u(o)$.

\emph{Step 2: the empty heights.} The preimages of the empty configuration are the height configurations with values in $U$. If $u\in U$ has a loop and an empty neighbour $u'\neq u$, then every independent set of sites can be given the height $u'$ inside a background of height $u$, which gives at least $2^{L^2/2}$ preimages. If $u$ has no loop but two distinct empty neighbours $u',u''$, then $h\equiv u$ on $\Lambda_0$ and $h$ arbitrary in $\{u',u''\}$ on $\Lambda_1$ give $2^{L^2/2}$ preimages. Hence every connected component of the graph induced on $U$ is a single vertex with a loop, a single edge between two unlooped vertices, or a single unlooped vertex. Let $a$ and $b$ be the numbers of components of the first two kinds. A height configuration with values in $U$ is constant on a looped vertex or alternates along an edge, so $\kappa=a+2b$.

\emph{Step 3: counting.} Let $x_v$ be the configuration with a single occupied site $v$. Since $T_L\setminus\{v\}$ is connected, a preimage restricts there to a configuration with values in one component of $U$, constant or alternating; the four neighbours of $v$ lie in the same sublattice and so carry a common height $u^*$, and $h(v)$ must be the occupied height matched to $u^*$. Hence the number of preimages of $x_v$ is the number of matched heights among the looped vertices and the vertices of edge components, which is at most $a+2b$, with equality only if all of them are matched. By (2), all of them are matched. Now the preimages of $x_\Lambda$ are constant on $\Lambda_0$ and on $\Lambda_1$, with values $o$ and $u(o)$, so their number is $|O|$, which is $a+2b$ plus the number of matched unlooped isolated empty heights. By (2), there are none. An unmatched unlooped isolated empty height has no neighbour at all and cannot occur in any configuration, which reducedness excludes.

\emph{Conclusion.} Every component of $(\cS,\sim)$ is therefore either a looped empty height $u$ together with its occupied partner, which is the binary alphabet, or an edge $u_1\sim u_2$ of empty heights together with the partners $o_1\sim u_1$ and $o_2\sim u_2$, which is the path $o_1-u_1-u_2-o_2$, that is, $A_4$ with occupied heights at the ends. There are no other adjacencies, since occupied heights are not adjacent to each other and are adjacent only to their partners.
\end{proof}

Uniformity in Definition~\ref{def:encoding}(2) matters only for disconnected alphabets. Suppose instead that, for every even $L$, every hard-square configuration on $T_L$ is the image of at least one and at most $K$ height configurations, with $K$ independent of $L$. The arguments of Steps~1 and~2 use only this bound, with $K$ in place of $\sup_M\kappa_M$. Since occupied heights are adjacent only to their partners, a connected reduced alphabet is then one component of the graph induced on $U$ together with the occupied heights matched to it. Besides the binary alphabet and $A_4$, this leaves alphabets without occupied heights, which miss every nonempty configuration; the path $o-u$ with $u$ unlooped, which has no configuration over the empty one; and the path $o-u_1-u_2$ of unlooped empty heights with $u_2$ unmatched. In the last case, let the only occupied sites be nonadjacent sites $v\in\Lambda_0$ and $w\in\Lambda_1$. Their neighbours carry $u_1$, and on the connected set $T_L\setminus\{v,w\}$ the heights alternate between $u_1$ and $u_2$, so they are constant on each sublattice. Hence $u_1$ is forced on both sublattices, which is impossible because $u_1$ has no loop. Thus a connected reduced alphabet with bounded multiplicity is the binary alphabet or $A_4$, and its multiplicity is uniform. A copy of $o-u_1-u_2$ next to a copy of the binary alphabet shows that uniformity is needed in the disconnected case.

\begin{remark}[The lift to $A_4$]\label{rem:lift}
The two-to-one correspondence of Example~\ref{ex:encodings}(2) is known. In the language of graph homomorphisms, the path $1-2-3-4$ is the universal cover of the graph of the binary alphabet, in the sense of \cite[Section~5]{ChandgotiaMarcus18}. What Proposition~\ref{prop:classification} adds is that there is no other case. It does not assume that the occupation map is a covering, since its hypothesis bounds only the number of height configurations over each hard-square configuration. In the literature on restricted solid-on-solid models, graphs with loops whose adjacency matrix has largest eigenvalue at most $2$ are viewed as simple graphs modulo a $\Z/2$ symmetry \cite[Section~6]{WarnaarNienhuis93}; the binary alphabet, whose adjacency matrix has largest eigenvalue $(1+\sqrt5)/2$, is in the same way the path $1-2-3-4$ modulo the reflection $\sigma$.
\end{remark}

\subsection{Presentations of hard squares}

A face weight $W$ on a height alphabet with occupation assigns to a height configuration $h$ on $T_L$ the weight $\prod W(h|_F)$ over the $L^2$ faces $F$ of $T_L$. The face with top-left corner $v$ has top row $h(v),h(v+e_1)$ and bottom row $h(v-e_2),h(v+e_1-e_2)$.

\begin{definition}\label{def:presentation}
A face weight $W$ \emph{presents hard squares} with activity $z\neq0$ if there is $c\neq0$ such that, for every even $L$ and every height configuration $h$ on $T_L$,
\[
 \prod_{F}W(h|_F)=c^{L^2}\,z^{|n\circ h|},
\]
where $|x|$ denotes the number of occupied sites of $x$.
\end{definition}

Thus a presentation reproduces hard squares configuration by configuration, at the level of heights. Activity zero is excluded, since otherwise the face weight equal to $1$ on the empty face and to $0$ elsewhere would present hard squares. On the binary alphabet the natural presentations are the following.

\begin{definition}\label{def:pure-hs-face}
A \emph{pure hard-square face} is a binary face weight of the form
\begin{equation}\label{eq:pure-hs-face}
 Q\fc abcd=A^aB^bC^cD^d\quad\text{on admissible faces},
\end{equation}
with corner factors $A,B,C,D\in\C^\times$. Its \emph{activity} is $z=ABCD$.
\end{definition}

There are exactly seven admissible binary faces, which we label
\begin{equation}\label{eq:seven-labels}
\begin{aligned}
 &w_0=\fc0000,\quad w_1=\fc0001,\quad w_2=\fc0010,\quad w_3=\fc0100,\\
 &w_4=\fc0110,\quad w_5=\fc1000,\quad w_6=\fc1001,
\end{aligned}
\end{equation}
and we write a binary face weight as $W=(w_0,\dots,w_6)$. A pure hard-square face is $Q=(1,D,C,B,BC,A,AD)$. The four corner factors express the freedom of distributing the activity of a site over the four faces containing it. By Proposition~\ref{prop:presentations}(1) below, a single binary face weight has no other freedom, since a gauge transformation of a pure hard-square face by factors on vertices or edges that cancel on every torus is again a multiple of a pure hard-square face. Face weights that alternate in a checkerboard pattern are face weights on $A_4$, those that alternate only by rows or only by columns are treated after Proposition~\ref{prop:presentations}, and distributions that are not periodic are not face weights in our sense.

Faces of $A_4$ are best described by occupations. Adjacent heights of $A_4$ have opposite parity, so in an admissible face $\fc abcd$ the heights $a,d$ have one parity and $b,c$ the other. We call the parity $\tau\in\{0,1\}$ of $a$ the \emph{type} of the face. Let $\phi_1(1)=1$, $\phi_1(0)=3$, $\phi_0(1)=4$, $\phi_0(0)=2$, so that $\phi_\tau$ maps an occupation to the height of parity $\tau$ carrying it. For a face weight $W$ on $A_4$ and $\tau\in\{0,1\}$ define the binary face weights
\begin{equation}\label{eq:types}
 W^\tau\fc xyuw=W\fc{\phi_\tau(x)}{\phi_{1-\tau}(y)}{\phi_{1-\tau}(u)}{\phi_\tau(w)} .
\end{equation}
As in Example~\ref{ex:encodings}, an $A_4$ face of type $\tau$ is admissible if and only if its occupations form an admissible binary face, so $W\mapsto(W^0,W^1)$ identifies face weights on $A_4$ with pairs of binary face weights.

\begin{proposition}[Presentations]\label{prop:presentations}
\begin{enumerate}
\item A binary face weight presents hard squares if and only if it is a nonzero multiple of a pure hard-square face.
\item If a face weight $W$ on $A_4$ presents hard squares with activity $z$, then $W^\tau=s_\tau Q^\tau$ with $s_\tau\neq0$ and $Q^\tau$ a pure hard-square face with corner factors $A_\tau,B_\tau,C_\tau,D_\tau$, for $\tau=0,1$, and $A_\tau B_{1-\tau}C_{1-\tau}D_\tau=z$. Conversely, every such weight presents hard squares with activity $z$, with $c^2=s_0s_1$.
\end{enumerate}
\end{proposition}

\begin{proof}
(1) Every admissible face occurs in some hard-square configuration, whose weight is nonzero, so $W$ has full support. A full-support binary face weight can be written uniquely as $W\fc abcd=w_0A^aB^bC^cD^dX^{ad}Y^{bc}$, with $A=w_5/w_0$, $B=w_3/w_0$, $C=w_2/w_0$, $D=w_1/w_0$, $X=w_6w_0/(w_5w_1)$ and $Y=w_4w_0/(w_3w_2)$. On a torus every site occupies each corner position of exactly one face, so a configuration $x$ receives the weight $w_0^{L^2}(ABCD)^{|x|}X^{N_\searrow(x)}Y^{N_\swarrow(x)}$, where $N_\searrow(x)$ and $N_\swarrow(x)$ count the occupied pairs $\{v,v+e_1-e_2\}$ and $\{v,v-e_1-e_2\}$. Comparing the empty configuration, a single particle, and a single isolated diagonal pair of each kind gives $w_0^{L^2}=c^{L^2}$, $ABCD=z$ and $X=Y=1$. The converse is the same computation.

(2) The same argument applies type by type. A height configuration on an even torus singles out the sublattice of odd heights; a site of parity $\pi$ is the top-left corner of a face of type $\pi$, the top-right and bottom-left corner of faces of type $1-\pi$, and the bottom-right corner of a face of type $\pi$. Hence the weight of $h$ is
\[
 (w^{(0)}_0w^{(1)}_0)^{L^2/2}\prod_{\pi}\bigl(A_\pi B_{1-\pi}C_{1-\pi}D_\pi\bigr)^{N_\pi}\prod_\tau X_\tau^{N_{\searrow,\tau}}Y_\tau^{N_{\swarrow,\tau}},
\]
where $N_\pi$ counts the occupied sites of parity $\pi$, and $N_{\searrow,\tau}$ and $N_{\swarrow,\tau}$ count the faces of type $\tau$ whose diagonal $\{a,d\}$, respectively $\{b,c\}$, is occupied. Every admissible face of either type occurs in some height configuration, so both $W^\tau$ have full support, and isolating each factor by suitable configurations gives $X_\tau=Y_\tau=1$ and $A_\tau B_{1-\tau}C_{1-\tau}D_\tau=z$. The converse is again the same computation.
\end{proof}

We call a pair $(s_0Q^0,s_1Q^1)$ as in Proposition~\ref{prop:presentations}(2) a \emph{staggered pure hard-square face}. In particular, every presentation of hard squares on either alphabet has full support. Through the lift of Example~\ref{ex:encodings}(2), face weights on $A_4$ are binary face weights that alternate between two types in a checkerboard pattern, so the staggered pure hard-square faces are the presentations of this kind. Weights that alternate only by rows or only by columns cannot be read from the heights of a height alphabet, whose adjacency rule is the same in both directions. On the binary alphabet the proof of (2) still applies to them, with the type of a face given by the parity of its top row, since a particle in a row of parity $k$ sits at the top corners of faces of type $k$ and at the bottom corners of faces of type $1-k$. Hence a pair $(W^0,W^1)$ of binary face weights, with $W^k$ on the faces whose top row has parity $k$, reproduces hard squares configuration by configuration exactly when each $W^k$ is a nonzero multiple of a pure hard-square face, with $A_kB_kC_{1-k}D_{1-k}=z$; columns are treated in the same way. By the remark after Theorem~\ref{thm:main}, no analytic Yang--Baxter family passes through either of them. This concerns one family through $W^0$ or through $W^1$; constructions with different families on rows of different parity are not covered.

Definition~\ref{def:presentation} asks for exactness configuration by configuration. It suffices to ask that the height weights reproduce hard squares after summing over each fibre of the occupation map, that is, that on each torus the total weight of the height configurations over a hard-square configuration is its hard-square weight.

\begin{proposition}[Fibre sums]\label{prop:fibre-sums}
Let $W$ be a face weight on $A_4$ and $z\neq0$. Suppose that for some even $L\ge6$ there is $c_L\neq0$ such that
\[
 \sum_{h\colon n\circ h=x}\ \prod_F W(h|_F)=c_L\,z^{|x|}
\]
for every hard-square configuration $x$ on $T_L$. Then $W$ is a staggered pure hard-square face of activity $z$; in particular it presents hard squares with activity $z$.
\end{proposition}

\begin{proof}
Let $\Lambda_0,\Lambda_1$ be the sublattices of $T_L$. By Example~\ref{ex:encodings}(2), the fibre over $x$ consists of the configuration $h_x$ with odd heights on $\Lambda_0$ and of $\sigma\circ h_x$, with $\sigma$ as in Example~\ref{ex:encodings}(2). A face of $h_x$ has type $1$ if its top-left site lies in $\Lambda_0$ and type $0$ otherwise, and $\sigma$ exchanges the types. For the empty configuration the hypothesis reads $2(w^{(0)}_0w^{(1)}_0)^{L^2/2}=c_L\neq0$, where $W^\tau=(w^{(\tau)}_0,\dots,w^{(\tau)}_6)$. We may therefore normalize $w^{(0)}_0=w^{(1)}_0=1$; then $c_L=2$, and only faces that meet occupied sites contribute. For $\pi\in\{0,1\}$ put
\[
 a_\pi=w^{(\pi)}_5\,w^{(1-\pi)}_3\,w^{(1-\pi)}_2\,w^{(\pi)}_1 .
\]
As in the proof of Proposition~\ref{prop:presentations}(2), a single particle on $\Lambda_0$ has weight $a_1$ in $h_x$ and $a_0$ in $\sigma\circ h_x$, and two particles on $\Lambda_0$ whose sets of faces are disjoint have weights $a_1^2$ and $a_0^2$. Hence
\[
 a_0+a_1=2z,\qquad a_0^2+a_1^2=2z^2,
\]
so $(a_0-a_1)^2=2(a_0^2+a_1^2)-(a_0+a_1)^2=0$ and $a_0=a_1=z\neq0$. In particular $w^{(\tau)}_k\neq0$ for $k=1,2,3,5$ and $\tau=0,1$, and we may put
\[
 X_\tau=\frac{w^{(\tau)}_6}{w^{(\tau)}_5w^{(\tau)}_1},\qquad Y_\tau=\frac{w^{(\tau)}_4}{w^{(\tau)}_3w^{(\tau)}_2}.
\]
The two occupied corners on the diagonal $\{a,d\}$ of a face with top-left site in $\Lambda_0$ have weights $z^2X_1$ in $h_x$ and $z^2X_0$ in $\sigma\circ h_x$; two such pairs with disjoint sets of faces have weights $z^4X_1^2$ and $z^4X_0^2$. Hence $X_0+X_1=2=X_0^2+X_1^2$, and $X_0=X_1=1$ by the same identity. The occupied corners on the diagonal $\{b,c\}$ of a face with top-left site in $\Lambda_1$ give $z^2Y_0$ and $z^2Y_1$ in the same way, and $Y_0=Y_1=1$. On $T_L$ with $L\ge6$ all these configurations exist, since the second particle, or the second pair, can be taken as the translate of the first by $3e_1-e_2$. The bound $L\ge6$ is needed because the faces that meet a diagonal pair have their top-left corners in a $3\times3$ square, and on $T_4$ two diagonal pairs of the same kind and type always share a face, although two particles on $\Lambda_0$ can be separated by $2e_1+2e_2$. Thus $W^\tau=(1,D_\tau,C_\tau,B_\tau,B_\tau C_\tau,A_\tau,A_\tau D_\tau)$ with $A_\tau=w^{(\tau)}_5$, $B_\tau=w^{(\tau)}_3$, $C_\tau=w^{(\tau)}_2$, $D_\tau=w^{(\tau)}_1$, and $A_\pi B_{1-\pi}C_{1-\pi}D_\pi=a_\pi=z$. Proposition~\ref{prop:presentations}(2) concludes.
\end{proof}

On the binary alphabet the fibres are single points, and the proof of Proposition~\ref{prop:presentations}(1) gives the same conclusion under the hypothesis of Proposition~\ref{prop:fibre-sums}.

\begin{remark}[Why encodings without hidden heights]\label{rem:why-class}
Among height alphabets, which carry the hard-core rule in their adjacency rule, the class of Definition~\ref{def:encoding} is not an ansatz, for the following reasons.
\begin{enumerate}
\item \emph{Transfer matrices.} On a cylinder of even circumference $M$, a row of heights of $A_4$ is a hard-square row together with the parity of its first height, and consecutive rows have opposite parities. Let $C_M$ be the compatibility matrix of hard-square rows and $Z_M$ the diagonal matrix with entries $z^{|x|}$, so that $C_MZ_M$ is the hard-square row transfer matrix. For a staggered pure hard-square face $(s_0Q^0,s_1Q^1)$ and $\pi\in\{0,1\}$, let $T_\pi$ be the transfer matrix onto rows of parity $\pi$, whose entry $(x,y)$ is the weight of the strip of $M$ faces with bottom row $x$ and top row $y$, where the first height of $y$ has parity $\pi$. Then $T_\pi=s\,G_\pi C_MG'_\pi$, with $s=(s_0s_1)^{M/2}$, where the diagonal matrices $G_\pi$ and $G'_\pi$ collect the corner factors that the faces of the strip give to the occupied sites of the bottom row and of the top row, respectively. The relation $A_\tau B_{1-\tau}C_{1-\tau}D_\tau=z$ says exactly that $G'_1G_0=G'_0G_1=Z_M$. The diagonal matrices $G_\pi$ are invertible, since the corner factors are nonzero, so $T_1T_0=s^2G_1(C_MZ_M)^2G_1^{-1}$ and $T_0T_1=s^2G_0(C_MZ_M)^2G_0^{-1}$. The row transfer matrix of $A_4$ acts on rows of both parities, mapping rows of parity $1-\pi$ to rows of parity $\pi$ by $T_\pi$, so its square is the direct sum of $T_1T_0$ and $T_0T_1$. A commuting family of row transfer matrices through a presentation on $A_4$ would thus contain a matrix whose square is, up to a scalar and a diagonal similarity, two copies of the square of the hard-square transfer matrix itself. This concerns families that come from face weights as in Definition~\ref{def:analytic-family}; Theorem~\ref{thmB} does not exclude that $C_MZ_M$ commutes with a family of transfer matrices of another origin. With hidden heights, the transfer matrix acts on a larger space of rows, and hard squares is recovered only after summing over the hidden heights.
\item \emph{Positive cases.} The class contains integrable models, such as hard hexagons on $A_4$ and Baxter's diagonal-interaction family on the binary alphabet (Section~\ref{sec:sharpness}). The obstruction of Theorem~\ref{thm:main} is therefore not an artefact of the class.
\item \emph{Maximality.} By Proposition~\ref{prop:classification} and the paragraph after it, a connected height alphabet that satisfies Definition~\ref{def:encoding}(1), in which every hard-square configuration on an even torus is an image, and that is neither the binary alphabet nor $A_4$, has unboundedly many height configurations over some hard-square configuration. In fact, on every torus $T_L$ with $L\ge4$ even, it has at least $2^{L^2/2}$ height configurations over the empty configuration or over the configuration that occupies one sublattice, because for connected alphabets the paragraph after Proposition~\ref{prop:classification} uses only these two fibres, through Steps~1 and~2, and the fact that every hard-square configuration on $T_L$ is an image. Within height alphabets, the boundary between Theorem~\ref{thm:main} and the open case thus does not depend on the bound allowed in Definition~\ref{def:encoding}(2).
\end{enumerate}
\end{remark}

\subsection{The Yang--Baxter equation on $A_4$ in occupation variables}

Through \eqref{eq:types}, the face Yang--Baxter equation on $A_4$ becomes a system for pairs of binary face weights.

\begin{lemma}[Staggered form]\label{lem:staggered}
Face weights $P,Q,R$ on $A_4$ satisfy \eqref{eq:FYB-rho} if and only if, for $\alpha\in\{0,1\}$ and all binary boundary data $(a,b,c,d,e,f)$,
\begin{equation}\label{eq:staggered}
 \sum_{g\in\{0,1\}}P^{1-\alpha}\fc fgab\,Q^{\alpha}\fc edfg\,R^{1-\alpha}\fc dcgb
 \;=\;\rho\sum_{g\in\{0,1\}}R^{\alpha}\fc egfa\,Q^{1-\alpha}\fc gcab\,P^{\alpha}\fc edgc .
\end{equation}
\end{lemma}

\begin{proof}
A term of either side of \eqref{eq:FYB-rho} is nonzero only if all six sides $ab,bc,cd,de,ef,fa$ of the hexagon join adjacent heights. Otherwise both sides vanish, so only such boundary data matter. For them the parities alternate around the hexagon. Let $\alpha$ be the parity of $a$, so that $a,c,e$ have parity $\alpha$ and $b,d,f$ parity $1-\alpha$. On the left of \eqref{eq:FYB-rho} the interior height $g$ is adjacent to $b,d,f$ and has parity $\alpha$; on the right it is adjacent to $a,c,e$ and has parity $1-\alpha$. The top-left corners of the faces on the left are $f,e,d$, of parities $1-\alpha,\alpha,1-\alpha$, and on the right $e,g,e$, of parities $\alpha,1-\alpha,\alpha$. Writing every height as $\phi_\pi$ of its occupation, \eqref{eq:FYB-rho} becomes \eqref{eq:staggered}, and since the $\phi_\pi$ are bijections the two systems are equivalent.
\end{proof}

In particular, the binary equation \eqref{eq:FYB} is the special case of \eqref{eq:staggered} in which $P^0=P^1$, $Q^0=Q^1$ and $R^0=R^1$, so the results on $A_4$ contain those on the binary alphabet for unstaggered weights.

\begin{example}[The critical restricted solid-on-solid model]\label{ex:rsos-calibration}
Let $\lambda=\pi/5$ and $S_h=\sin(h\lambda)$. For admissible faces of $A_4$ put
\[
 W_u\fc abcd=\sin(\lambda-u)\,\delta_{b,c}+\sin u\,\delta_{a,d}\,\frac{\sqrt{S_bS_c}}{S_a},
\]
the face form of the Temperley--Lieb Baxterization $\sin(\lambda-u)\,1+\sin u\,e$ in the path representation of $A_4$. Then $(P,Q,R)=(W_u,W_{u+v},W_v)$ satisfy \eqref{eq:FYB} for all $u,v$; this is the standard Yang--Baxter equation of the critical restricted solid-on-solid model, here in the convention of Definition~\ref{def:fyb} (Appendix~\ref{app:computations}). The check does not distinguish $(W_u,W_{u+v},W_v)$ from $(W_v,W_{u+v},W_u)$, and nothing depends on it, since a family that satisfies \eqref{eq:FYB-rho} for $(W_{qr},W_{pr},W_{pq})$ becomes an analytic Yang--Baxter family after $W_{uv}\mapsto W_{vu}$, passing through the same faces. The weights $W_u$ have full support for generic $u$, but they do not present hard squares. Transported to occupations by \eqref{eq:types}, they are the same for both types, so by the remark after Lemma~\ref{lem:staggered} the binary weights $W^0_{v-u}$ form an analytic Yang--Baxter family on the binary alphabet. As Theorem~\ref{thm:main} requires, they never present hard squares. With $s=\sin u$ and $t=\sin(\lambda-u)$, full support needs $st\neq0$, and the ratios $X$ and $Y$ of the proof of Proposition~\ref{prop:presentations}(1) equal $1$ only for $t=-s/\varphi$ and for $t=-\varphi s$ respectively, where $\varphi$ is the golden ratio. At $u=-\lambda=-\pi/5$ they satisfy $w_6=0$ and $w_4w_0=w_3w_2$. This point lies outside the interval $0<u<\lambda$ on which all the weights $W_u$ are positive, and some face weights are negative there, although every configuration weight is positive. In Baxter's parametrization of the diagonal-interaction model, hard hexagons is likewise reached at $u=-\pi/5$, the endpoint of the range $-\pi/5<u<0$ of his regimes I and II \cite[Eqs.~(41)--(44)]{Baxter80}. By the computation in the proof of Proposition~\ref{prop:presentations}(1), they give the weight $w_0^{L^2}z^{|x|}$ to the hard-square configurations $x$ with no occupied pair $\{v,v+e_1-e_2\}$ and $0$ to the others, with $z=\varphi^5=(11+5\sqrt5)/2$. This is the hard-hexagon model at its critical activity \cite{Baxter80}, a positive control for the convention and the occupation map. Away from criticality, the hard-hexagon model is a specialization of the off-critical family on the same alphabet \cite{ABF84}.
\end{example}

\section{The obstruction}\label{sec:support}

Boundary data are written $(a,b,c,d,e,f)$ as in \eqref{eq:FYB}. For each boundary datum, \eqref{eq:FYB} and \eqref{eq:staggered} are bilinear in the weights of $P$ and $R$, with coefficients that are monomials in the corner factors of the middle face. We write $P=(p_0,\dots,p_6)$ and $R=(r_0,\dots,r_6)$ in the labelling \eqref{eq:seven-labels}, and $P^\tau=(p^{(\tau)}_0,\dots,p^{(\tau)}_6)$, $R^\tau=(r^{(\tau)}_0,\dots,r^{(\tau)}_6)$ for the two types on $A_4$.

\subsection{The binary certificate}

Let $Q$ be a pure hard-square face \eqref{eq:pure-hs-face} with $A,B,C,D\neq0$. Enumerating the $2^6$ boundary data, exactly eighteen of the equations \eqref{eq:FYB} are not identically zero. We use seven of them:
\begin{align}
 (0,0,1,0,0,0):&\quad p_0r_3+Dp_3r_4-Bp_1r_0=0, \label{eq:b1}\\
 (0,0,1,0,1,0):&\quad Ap_0r_3+ADp_3r_4-Bp_6r_5=0, \label{eq:b2}\\
 (0,0,0,1,0,0):&\quad Bp_0r_5-p_3r_0-Ap_4r_3=0, \label{eq:b3}\\
 (0,1,0,1,0,0):&\quad Bp_1r_6-Dp_3r_0-ADp_4r_3=0, \label{eq:b4}\\
 (0,0,0,1,0,1):&\quad BCp_5r_5-p_3r_2-Ap_4r_4=0, \label{eq:b5}\\
 (0,1,0,1,0,1):&\quad BCp_6r_6-Dp_3r_2-ADp_4r_4=0, \label{eq:b6}\\
 (0,0,0,0,1,0):&\quad ADp_3r_2+Ap_0r_0-p_5r_5=0. \label{eq:b7}
\end{align}
For instance, for $(0,0,0,0,1,0)$ the left and right sides of \eqref{eq:FYB} are
\begin{gather*}
 P\fc0000Q\fc1000R\fc0000+P\fc0100Q\fc1001R\fc0010=Ap_0r_0+ADp_3r_2,\\
 R\fc1000Q\fc0000P\fc1000=r_5p_5,
\end{gather*}
respectively, since on the right the term with $g=1$ contains an inadmissible face. Each equation of \eqref{eq:FYB} has at most four terms, two on each side, and in each of \eqref{eq:b1}--\eqref{eq:b7} the exclusion rule leaves only three.

\begin{theorem}[Binary certificate]\label{thm:binary-certificate}
If $P$ and $R$ satisfy \eqref{eq:FYB} with middle face $Q$, then $r_5p_3r_2=0$. More precisely, put
\[
 E_1=p_6r_5-Ap_1r_0,\quad E_2=p_1r_6-Dp_0r_5,\quad E_3=p_6r_6-Dp_5r_5,\quad E_4=p_5r_5-Ap_0r_0-ADp_3r_2 .
\]
Then $E_1,E_2,E_3,E_4$ vanish on every solution, and
\begin{equation}\label{eq:binary-certificate}
 AD^2\,r_5p_3r_2=r_6E_1+Ar_0E_2-r_5E_3-Dr_5E_4
\end{equation}
identically in the fourteen auxiliary weights.
\end{theorem}

\begin{proof}
Subtracting \eqref{eq:b2} from $A$ times \eqref{eq:b1} gives $BE_1=0$; subtracting $D$ times \eqref{eq:b3} from \eqref{eq:b4} gives $BE_2=0$; subtracting $D$ times \eqref{eq:b5} from \eqref{eq:b6} gives $BCE_3=0$; and \eqref{eq:b7} is $-E_4=0$. Since $B,C\neq0$, all $E_i$ vanish. Expanding, the right side of \eqref{eq:binary-certificate} is
\[
 r_6p_6r_5-Ar_6p_1r_0+Ar_0p_1r_6-ADr_0p_0r_5-r_5p_6r_6+Dr_5p_5r_5-Dr_5p_5r_5+ADr_5p_0r_0+AD^2r_5p_3r_2 ,
\]
which equals the left side. Since $A,D\neq0$, the claim follows.
\end{proof}

\begin{remark}[Orientations and scalar factors]\label{rem:orientations}
The dihedral group of the square permutes the seven admissible faces, preserves full support, and maps pure hard-square faces to pure hard-square faces with permuted corner factors, so Theorem~\ref{thm:binary-certificate} applies to the three faces in arbitrary fixed orientations. With \eqref{eq:FYB-rho} in place of \eqref{eq:FYB}, the boundary data $(0,0,1,0,0,0)$, $(0,0,1,0,1,0)$, $(0,0,0,0,0,1)$, $(0,1,0,0,0,1)$ and $(0,1,0,0,1,0)$ yield, for $\rho\neq0$, the vanishing of
\[
 F_1=p_6r_5-Ap_1r_0,\qquad F_2=p_6r_1-Dp_5r_0,\qquad F_3=Ap_1r_1-\rho Dp_5r_5,
\]
and $r_0F_3+r_1F_1-r_5F_2=(1-\rho)D\,r_0p_5r_5$. Thus $\rho=1$ as soon as $r_0p_5r_5\neq0$, and a scalar factor reduces to \eqref{eq:FYB}.

The positions of the three faces can be permuted as well. For a face weight $W$ put $W^\theta\fc abcd=W\fc bdac$, the face weight rotated by a quarter turn, so that $W^{\theta^2}\fc abcd=W\fc dcba$. Substituting these definitions, the equation \eqref{eq:FYB-rho} for $(R^{\theta^2},P^\theta,Q^\theta)$ with boundary datum $(a,b,c,d,e,f)$ is the equation for $(P,Q,R)$ with boundary datum $(e,f,a,b,c,d)$, with the same $\rho$, on every height alphabet; applying this twice brings $R$ to the middle. With the reduction to $\rho=1$ above, Theorem~\ref{thm:binary-certificate} therefore shows that if any one of the three faces is a pure hard-square face, the other two do not both have full support.
\end{remark}

\subsection{The certificate on the hard-hexagon alphabet}

On $A_4$ the middle face is a staggered pure hard-square face $(s_0Q^0,s_1Q^1)$, with corner factors $A_\tau,B_\tau,C_\tau,D_\tau$ for $Q^\tau$. In the equations of \eqref{eq:staggered} with a given $\alpha$, the left side contains $Q^\alpha$ and the right side $Q^{1-\alpha}$; dividing them by $s_\alpha$ we may take $s_0=s_1=1$, at the cost of replacing $\rho$ by
\[
 \rho_\alpha=\rho\,s_{1-\alpha}/s_\alpha\qquad\text{in the equations with parity }\alpha .
\]
Among the $128$ equations of \eqref{eq:staggered}, exactly $36$ are not identically zero. We use eight of them, indexed by $(\alpha;a,b,c,d,e,f)$:
\begin{align}
 (0;0,0,1,0,0,0):&\quad p^{(1)}_0r^{(1)}_3+D_0p^{(1)}_3r^{(1)}_4-\rho_0B_1\,p^{(0)}_1r^{(0)}_0=0, \label{eq:s1}\\
 (0;0,0,1,0,1,0):&\quad A_0p^{(1)}_0r^{(1)}_3+A_0D_0p^{(1)}_3r^{(1)}_4-\rho_0B_1\,p^{(0)}_6r^{(0)}_5=0, \label{eq:s2}\\
 (1;0,0,0,1,0,0):&\quad B_1\,p^{(0)}_0r^{(0)}_5-\rho_1\bigl(p^{(1)}_3r^{(1)}_0+A_0p^{(1)}_4r^{(1)}_3\bigr)=0, \label{eq:s3}\\
 (1;0,1,0,1,0,0):&\quad B_1\,p^{(0)}_1r^{(0)}_6-\rho_1D_0\bigl(p^{(1)}_3r^{(1)}_0+A_0p^{(1)}_4r^{(1)}_3\bigr)=0, \label{eq:s4}\\
 (1;0,0,0,1,0,1):&\quad B_1C_1\,p^{(0)}_5r^{(0)}_5-\rho_1\bigl(p^{(1)}_3r^{(1)}_2+A_0p^{(1)}_4r^{(1)}_4\bigr)=0, \label{eq:s5}\\
 (1;0,1,0,1,0,1):&\quad B_1C_1\,p^{(0)}_6r^{(0)}_6-\rho_1D_0\bigl(p^{(1)}_3r^{(1)}_2+A_0p^{(1)}_4r^{(1)}_4\bigr)=0, \label{eq:s6}\\
 (0;0,0,0,0,1,0):&\quad A_0p^{(1)}_0r^{(1)}_0+A_0D_0p^{(1)}_3r^{(1)}_2-\rho_0\,p^{(0)}_5r^{(0)}_5=0, \label{eq:s7}\\
 (0;0,0,0,0,0,0):&\quad p^{(1)}_0r^{(1)}_0+D_0p^{(1)}_3r^{(1)}_2-\rho_0\bigl(p^{(0)}_0r^{(0)}_0+A_1p^{(0)}_2r^{(0)}_3\bigr)=0. \label{eq:s8}
\end{align}
The weights of the two types are interlocked; the certificate below eliminates those of type $1$.

\begin{theorem}[Certificate on $A_4$]\label{thm:a4-certificate}
Let $(s_0Q^0,s_1Q^1)$ be a staggered pure hard-square face with nonzero corner factors, let $P^\tau,R^\tau$ satisfy \eqref{eq:staggered} with a scalar factor $\rho\neq0$, and put $\rho_0=\rho\,s_1/s_0$. Then
\[
 r^{(0)}_5\,p^{(0)}_2\,r^{(0)}_3=0\qquad\text{and}\qquad r^{(1)}_5\,p^{(1)}_2\,r^{(1)}_3=0 .
\]
More precisely, with
\begin{gather*}
 G_1=p^{(0)}_6r^{(0)}_5-A_0p^{(0)}_1r^{(0)}_0,\qquad G_2=p^{(0)}_1r^{(0)}_6-D_0p^{(0)}_0r^{(0)}_5,\\
 G_3=p^{(0)}_6r^{(0)}_6-D_0p^{(0)}_5r^{(0)}_5,\qquad
 G_4=\rho_0p^{(0)}_5r^{(0)}_5-A_0p^{(1)}_0r^{(1)}_0-A_0D_0p^{(1)}_3r^{(1)}_2,\\
 G_5=p^{(1)}_0r^{(1)}_0+D_0p^{(1)}_3r^{(1)}_2-\rho_0p^{(0)}_0r^{(0)}_0-\rho_0A_1p^{(0)}_2r^{(0)}_3,
\end{gather*}
all $G_i$ vanish on every solution, and
\begin{equation}\label{eq:a4-certificate}
 \rho_0A_0A_1D_0\;r^{(0)}_5\,p^{(0)}_2\,r^{(0)}_3
 =\rho_0\bigl(r^{(0)}_6G_1+A_0r^{(0)}_0G_2-r^{(0)}_5G_3\bigr)-D_0r^{(0)}_5G_4-A_0D_0r^{(0)}_5G_5
\end{equation}
identically in the twenty-eight auxiliary weights.
\end{theorem}

\begin{proof}
Subtracting \eqref{eq:s2} from $A_0$ times \eqref{eq:s1} gives $\rho_0B_1G_1=0$; subtracting $D_0$ times \eqref{eq:s3} from \eqref{eq:s4} gives $B_1G_2=0$, the terms with $\rho_1$ cancelling; subtracting $D_0$ times \eqref{eq:s5} from \eqref{eq:s6} gives $B_1C_1G_3=0$; and $G_4$ and $G_5$ are minus the left side of \eqref{eq:s7} and the left side of \eqref{eq:s8}. Since $\rho_0,B_1,C_1\neq0$, all $G_i$ vanish. For the identity, first note that, as in \eqref{eq:binary-certificate},
\[
 r^{(0)}_6G_1+A_0r^{(0)}_0G_2-r^{(0)}_5G_3=D_0r^{(0)}_5\bigl(p^{(0)}_5r^{(0)}_5-A_0p^{(0)}_0r^{(0)}_0\bigr).
\]
Next, $D_0r^{(0)}_5G_4+A_0D_0r^{(0)}_5G_5=\rho_0D_0r^{(0)}_5\bigl(p^{(0)}_5r^{(0)}_5-A_0p^{(0)}_0r^{(0)}_0-A_0A_1p^{(0)}_2r^{(0)}_3\bigr)$, because the terms of type $1$ cancel. Subtracting gives \eqref{eq:a4-certificate}. Since $\rho_0,A_0,A_1,D_0\neq0$, the first claim follows, and the second follows by exchanging the roles of the two types.
\end{proof}

The proof of the first claim uses $\rho_0\neq0$ and does not involve $\rho_1$, so Theorem~\ref{thm:a4-certificate} holds as well when the equations of parity $0$ and $1$ carry independent nonzero factors $\rho_0$ and $\rho_1$. When the weights are unstaggered, $\rho_0=\rho_1=\rho$, and the same computation gives $r_5p_2r_3=0$ for every $\rho\neq0$, a companion of Theorem~\ref{thm:binary-certificate} that does not need the reduction to $\rho=1$ of Remark~\ref{rem:orientations}. The exclusion rule enters in the same way in both certificates. Each equation of \eqref{eq:staggered} has at most four terms, and in each of \eqref{eq:s1}--\eqref{eq:s7} only three survive.

As in Remark~\ref{rem:orientations}, a symmetry of the square, acting on the corners of the faces of $A_4$, maps a staggered pure hard-square face to a staggered pure hard-square face with permuted corner factors. It exchanges the two types when it moves the top-left corner to the top-right or the bottom-left one, because adjacent corners carry heights of opposite parity. Since Theorem~\ref{thm:a4-certificate} concludes for both types, it applies to the three faces in arbitrary fixed orientations and, by the cyclic symmetry of Remark~\ref{rem:orientations}, in arbitrary positions. This matters because the calibration of Example~\ref{ex:rsos-calibration} does not fix the orientation. The weights $W_u$ are invariant under $b\leftrightarrow c$ and under $a\leftrightarrow d$, so they satisfy \eqref{eq:FYB} also when every face is read transposed.

\subsection{The main theorem}

\begin{theorem}\label{thm:main}
Let $(\cS,\sim,n)$ be the binary alphabet or $A_4$, and let $(W_{uv})$ be an analytic Yang--Baxter family on it (Definition~\ref{def:analytic-family}). Then no $W_{p_0r_0}$ presents hard squares with nonzero activity. The same holds if $W_{p_0r_0}$ is only required to reproduce hard squares after summing over the fibres of the occupation map, as in Proposition~\ref{prop:fibre-sums}.
\end{theorem}

\begin{proof}
Suppose $W_{p_0r_0}$ presents hard squares with activity $z\neq0$, or reproduces them after fibre sums. By Propositions~\ref{prop:presentations} and~\ref{prop:fibre-sums} it is a multiple of a pure hard-square face, respectively a staggered pure hard-square face, and in either case it has full support and nonzero corner factors. Put $P=W_{p_0q}$ and $R=W_{qr_0}$ for $q\in\mathcal D$. Each weight of $P$, respectively of $P^\tau$, is a holomorphic function of $q$ that equals the corresponding nonzero weight of $W_{p_0r_0}$ at $q=r_0$; likewise each weight of $R$ is nonzero at $q=p_0$. None of these finitely many functions vanishes identically on the connected domain $\mathcal D$, so for $q$ outside a discrete set all of them are nonzero.

On the binary alphabet, $W_{p_0r_0}=sQ$ with $s\neq0$ and $Q$ a pure hard-square face; since \eqref{eq:FYB-rho} is linear in the middle face, $P$ and $R$ satisfy it with middle face $Q$ and the same $\rho$. Remark~\ref{rem:orientations} gives $\rho(p_0,q,r_0)=1$ for such $q$, and Theorem~\ref{thm:binary-certificate} gives $r_5p_3r_2=0$, a contradiction. On $A_4$, Lemma~\ref{lem:staggered} turns the Yang--Baxter equation into \eqref{eq:staggered}, and Theorem~\ref{thm:a4-certificate} gives $r^{(0)}_5p^{(0)}_2r^{(0)}_3=0$, again a contradiction.
\end{proof}

Theorem~\ref{thmB} of the introduction is Theorem~\ref{thm:main}, and Theorem~\ref{thmA} is Proposition~\ref{prop:classification}. Combining them, we obtain the following.

\begin{corollary}[Reduced encodings]\label{cor:reduced}
No reduced height alphabet that encodes hard squares without hidden heights carries an analytic Yang--Baxter family passing through a presentation of hard squares with nonzero activity.
\end{corollary}

\begin{proof}
By Proposition~\ref{prop:classification}, the alphabet is a disjoint union of copies of the binary alphabet and of $A_4$. In a nonzero term of either side of \eqref{eq:FYB-rho} the three faces are admissible, so the seven heights of the term lie in one component. Hence the equations \eqref{eq:FYB-rho} on the union are those on the components, and the restrictions of the $W_{uv}$ to a component form an analytic Yang--Baxter family on it, with the same scalar factor. A height configuration on a torus takes values in one component, so a presentation of hard squares restricts to a presentation of hard squares on each component, with the same $c$ and $z$. Theorem~\ref{thm:main} concludes.
\end{proof}

Corollary~\ref{cor:reduced} concerns presentations. Its proof does not apply to the fibre-sum version, since on a fixed torus the components can compensate each other, and we do not claim that version for disjoint unions.

Theorem~\ref{thm:main} covers every nonzero activity, including $z=-1$, where all eigenvalues of the row-to-row transfer matrix with periodic rows, in the standard orientation, are roots of unity, as conjectured by Fendley, Schoutens and van Eerten \cite{FendleySchoutensvanEerten05} and proved by Jonsson through periodic rhombus tilings \cite{Jonsson06}; with free row ends this is observed for small widths, and in the diagonal orientation some eigenvalues are zero or roots of $x^3=4\cos^2(\pi m/N)$ \cite{Baxter11}. Whatever explains this structure, it is not an analytic Yang--Baxter family through a presentation on the binary alphabet or on $A_4$. The proof uses Propositions~\ref{prop:presentations} and~\ref{prop:fibre-sums} only to know that $W_{p_0r_0}$ is a nonzero multiple of a pure hard-square face, respectively a pair $(s_0Q^0,s_1Q^1)$ of pure hard-square faces with $s_0s_1\neq0$, with nonzero corner factors. The certificates of Theorems~\ref{thm:binary-certificate} and~\ref{thm:a4-certificate} do not use the relation between the corner factors and $z$. Theorem~\ref{thm:main} thus also excludes analytic families through such faces when they do not present hard squares. On $A_4$ these include the staggered faces with $A_0B_1C_1D_0\neq A_1B_0C_0D_1$. By the proof of Proposition~\ref{prop:presentations}(2), such a face gives the activity $A_\pi B_{1-\pi}C_{1-\pi}D_\pi$ to the occupied sites of height parity $\pi$, so on each height configuration it is hard squares with different activities on the two sublattices. A binary face weight cannot do this, since every site occupies each corner position of exactly one face.

\subsection{Sharpness}\label{sec:sharpness}

\emph{Diagonal interactions restore integrability.} The obstruction is specific to the pure hard-core rule. On $A_4$, the hard-hexagon model, which also excludes one diagonal, is a specialization of the off-critical restricted solid-on-solid family \cite{ABF84}. Its presentation has a vanishing diagonal weight, and the certificate does not apply. On the binary alphabet, the hard-square gas with diagonal couplings $M,L$ has the weights $\mathbf 1_{\mathrm{hc}}\,z^{(a+b+c+d)/4}e^{Mad+Lbc}$ and is integrable on Baxter's surface \cite{Baxter80,Baxter81,BaxterBook,NyckeesMila22}
\begin{equation}\label{eq:baxter-surface}
 z=\frac{(1-e^{-L})(1-e^{-M})}{e^{L+M}-e^{L}-e^{M}} .
\end{equation}
With $X=e^{M}$ and $Y=e^{L}$ it reads $(X-1)(Y-1)=zXY(XY-X-Y)$, and its closure is $\{F=0\}$, where $F(z,X,Y)=zXY(XY-X-Y)-(X-1)(Y-1)$. The pure hard-square model is the line $\{X=Y=1\}$. The hard-hexagon model, which excludes the diagonal $\{a,d\}$, is the line $\{X=0,\ Y=1\}$, the limit $M\to-\infty$ at $L=0$; it lies on $\{F=0\}$ for every $z$.

\begin{remark}[Baxter's surface]\label{rem:baxter-surface}
The surface $\{F=0\}$ meets the pure hard-square line only at $z=0$, and if $(z_j,X_j,Y_j)\in\{F=0\}$ and $(X_j,Y_j)\to(1,1)$, then $z_j\to0$. Indeed $F(z,1,1)=-z$, and near $(X,Y)=(1,1)$ one has $XY(XY-X-Y)\to-1$, so on the surface $z=(X-1)(Y-1)/\bigl(XY(XY-X-Y)\bigr)\to0$. This is noted after Eq.~(24) of \cite{Baxter80} and in \cite[Appendix~B]{TanakaTamuraKatsura12}. Theorem~\ref{thm:main} replaces this single known family by all analytic families on both alphabets.
\end{remark}

\emph{Support loss is sharp for a single equation.} Theorem~\ref{thm:binary-certificate} forces one vanishing auxiliary weight. In fact one zero never suffices, and the solutions with exactly two zeros are explicit.

\begin{theorem}[Near-full support]\label{thm:near-full}
Let $P,R$ satisfy \eqref{eq:FYB} with a pure hard-square middle face $Q$, $A,B,C,D\neq0$. Then at least two of the fourteen auxiliary weights vanish. If exactly two vanish, then either $r_2=r_3=0$ and, with free nonzero parameters $p_0,r_0,r_1,r_5$,
\begin{equation}\label{eq:branch-r}
\begin{gathered}
 p_1=\frac{Dp_0r_0}{r_1},\quad p_2=\frac{Cp_0r_1}{r_0},\quad p_3=\frac{Bp_0r_5}{r_0},\quad
 p_4=\frac{BCp_0r_1r_5}{r_0^2},\quad p_5=\frac{Ap_0r_0}{r_5},\\
 p_6=\frac{ADp_0r_0^2}{r_1r_5},\qquad r_4=\frac{r_0^3}{r_1r_5},\qquad r_6=\frac{r_1r_5}{r_0};
\end{gathered}
\end{equation}
or $p_2=p_3=0$ and, with free nonzero parameters $p_0,p_1,p_5,r_0$,
\begin{equation}\label{eq:branch-p}
\begin{gathered}
 p_4=\frac{p_0^3}{p_1p_5},\quad p_6=\frac{p_1p_5}{p_0},\quad r_1=\frac{Dp_0r_0}{p_1},\quad r_2=\frac{Cp_5r_0}{p_0},\quad r_3=\frac{Bp_1r_0}{p_0},\\
 r_4=\frac{BCp_1p_5r_0}{p_0^2},\qquad r_5=\frac{Ap_0r_0}{p_5},\qquad r_6=\frac{ADp_0^2r_0}{p_1p_5}.
\end{gathered}
\end{equation}
Conversely, \eqref{eq:branch-r} and \eqref{eq:branch-p} satisfy all eighteen equations.
\end{theorem}

\begin{proof}
Besides \eqref{eq:b1}--\eqref{eq:b7} we use the equations for $(1,0,0,1,0,0)$ and $(0,0,1,0,0,1)$:
\begin{equation}\label{eq:b8}
 Bp_2r_5-Cp_3r_1=0,\qquad Cp_5r_3-Bp_1r_2=0 .
\end{equation}
By Theorem~\ref{thm:binary-certificate}, one of $r_5,p_3,r_2$ vanishes. If $r_5=0$, then $E_1=0$ gives $p_1r_0=0$ and $E_3=0$ gives $p_6r_6=0$, so at least three weights vanish. If $p_3=0\neq r_5$, the first equation in \eqref{eq:b8} forces $p_2=0$. If $r_2=0$ and $p_3r_5\neq0$, the second equation gives $p_5r_3=0$; and if $p_5=0$, then $E_4=0$ gives $p_0r_0=0$, so three weights vanish. Hence at least two weights vanish, and if exactly two vanish the zero set is $\{p_2,p_3\}$ or $\{r_2,r_3\}$.

Suppose $r_2=r_3=0$ and all other weights are nonzero. Then \eqref{eq:b7} and \eqref{eq:b3} give $p_5$ and $p_3$ in terms of $p_0,r_0,r_5$; the equation for $(0,1,0,0,0,0)$, which reads $p_1r_1-Dp_0r_0-ADp_2r_3=0$, gives $p_1$; then $E_2=0$ gives $r_6$, and \eqref{eq:b1} gives $Dp_3r_4=Bp_1r_0$, hence $r_4$; the equations for $(0,0,0,0,0,1)$ and $(0,1,0,0,0,1)$, namely $Cp_5r_0-Ap_2r_4-p_0r_2=0$ and $Cp_6r_1-ADp_2r_4-Dp_0r_2=0$, give $p_2$ and $p_6$; finally \eqref{eq:b5} gives $p_4$. Each step solves a linear equation with a nonzero coefficient, so every solution on this support is of the form \eqref{eq:branch-r}. The branch $p_2=p_3=0$ is treated in the same way, and the converse is a direct substitution (Appendix~\ref{app:computations}).
\end{proof}

The toric families are single solutions of one Yang--Baxter equation and do not form spectral families. They lose the weights $r_2,r_3$ or $p_2,p_3$, which are nonzero in the middle face, and so cannot arise as $W_{p_0q}$, $W_{qr_0}$ for $q$ near $r_0$, $p_0$. Up to corner factors whose product is $1$, they are the triples $(Q,Q,\mathrm{Id})$ and $(\mathrm{Id},Q,Q)$, where $\mathrm{Id}$ is the face weight $\delta_{b,c}$. In \eqref{eq:branch-r}, with $\alpha=r_1/r_0$ and $\beta=r_5/r_0$, one has $R\fc abcd=r_0\beta^a\alpha^{-b}\beta^{-c}\alpha^d\delta_{b,c}$, and $P$ is $p_0$ times the pure hard-square face with corner factors $A/\beta$, $B\beta$, $C\alpha$, $D/\alpha$. In \eqref{eq:branch-p}, with $\alpha=p_1/p_0$ and $\beta=p_5/p_0$, one has $P\fc abcd=p_0\beta^a\alpha^{-b}\beta^{-c}\alpha^d\delta_{b,c}$, and $R$ is $r_0$ times the pure hard-square face with corner factors $A/\beta$, $B\alpha$, $C\beta$, $D/\alpha$. These are the values that a family with $W_{uu}$ proportional to $\mathrm{Id}$ would take at $q=r_0$ and at $q=p_0$.

\emph{Intertwiners outside a family.} The family hypothesis can be traded for full support of one auxiliary face, and the toric families are then the only solutions.

\begin{proposition}[Rigidity of the intertwiner]\label{prop:rigidity}
Let $Q$ be a pure hard-square face with $A,B,C,D\neq0$, and let $P,R$ be nonzero face weights on the binary alphabet that satisfy \eqref{eq:FYB-rho} with middle face $Q$ for some $\rho\neq0$. If $P$ has full support, then $\rho=1$ and $(P,R)$ is given by \eqref{eq:branch-r}. If $R$ has full support, then $\rho=1$ and $(P,R)$ is given by \eqref{eq:branch-p}.
\end{proposition}

\begin{proof}
By Remark~\ref{rem:orientations}, $F_1=p_6r_5-Ap_1r_0$ and $F_2=p_6r_1-Dp_5r_0$ vanish, and $r_0F_3+r_1F_1-r_5F_2=(1-\rho)D\,r_0p_5r_5$. We also use the equations \eqref{eq:FYB-rho} for seven boundary data:
\begin{align*}
 (0,0,0,0,0,0):&\quad Dp_3r_2+(1-\rho)p_0r_0-\rho Ap_2r_3=0,\\
 (0,0,0,0,1,0):&\quad ADp_3r_2+Ap_0r_0-\rho p_5r_5=0,\\
 (0,0,0,1,0,0):&\quad Bp_0r_5-\rho p_3r_0-\rho Ap_4r_3=0,\\
 (0,0,1,0,0,0):&\quad p_0r_3+Dp_3r_4-\rho Bp_1r_0=0,\\
 (0,1,0,0,0,0):&\quad p_1r_1-\rho Dp_0r_0-\rho ADp_2r_3=0,\\
 (0,1,0,1,0,0):&\quad Bp_1r_6-\rho Dp_3r_0-\rho ADp_4r_3=0,\\
 (1,0,0,1,0,0):&\quad Bp_2r_5-\rho Cp_3r_1=0 .
\end{align*}

Suppose that $P$ has full support. By $F_1=F_2=0$, the weights $r_0,r_1,r_5$ vanish together. If they vanish, the second, first, fourth and sixth equations give in turn $r_2=0$, $r_3=0$, $r_4=0$ and $r_6=0$, so $R=0$, which is excluded. Hence $r_0r_1r_5\neq0$, and $\rho=1$ since $r_0p_5r_5\neq0$. Theorem~\ref{thm:binary-certificate} gives $r_2=0$, because $r_5p_3\neq0$; then the first equation gives $r_3=0$, and the sixth and fourth give $Bp_1r_6=Dp_3r_0\neq0$ and $Dp_3r_4=Bp_1r_0\neq0$. Thus exactly $r_2$ and $r_3$ vanish, and Theorem~\ref{thm:near-full} gives \eqref{eq:branch-r}.

Suppose that $R$ has full support. By $F_1=F_2=0$, the weights $p_1,p_5,p_6$ vanish together. If they vanish, the sixth equation gives $Ap_4r_3=-p_3r_0$, so the third gives $Bp_0r_5=0$ and $p_0=0$; then the fifth, second and sixth equations give in turn $p_2=0$, $p_3=0$ and $p_4=0$, so $P=0$. Hence $p_1p_5p_6\neq0$, and $\rho=1$. Theorem~\ref{thm:binary-certificate} gives $p_3=0$; then the seventh equation gives $p_2=0$, and the second and third give $Ap_0r_0=p_5r_5\neq0$ and $Ap_4r_3=Bp_0r_5\neq0$. Thus exactly $p_2$ and $p_3$ vanish, and Theorem~\ref{thm:near-full} gives \eqref{eq:branch-p}.
\end{proof}

In the train argument, $P$ and $Q$ are faces of consecutive rows and $R$ intertwines them (Section~\ref{sec:prelim}). Proposition~\ref{prop:rigidity} uses no family. On the binary alphabet, a face weight that intertwines the hard-square face with a row face of full support is, up to a scalar and the factors $\alpha,\beta$ above, the face $\mathrm{Id}$, and that row face is a multiple of a pure hard-square face; and an intertwiner of full support pairs the hard-square face only with a row face that is, in the same sense, the face $\mathrm{Id}$. By the cyclic symmetry of Remark~\ref{rem:orientations}, the proposition applies with the pure hard-square face in any of the three positions. If one of the two other faces has full support, the triple is obtained from \eqref{eq:branch-r} or \eqref{eq:branch-p} by those symmetries. In particular, row faces of full support that are intertwined with the hard-square face by face weights satisfying \eqref{eq:FYB-rho}, whether or not these belong to a family, are multiples of pure hard-square faces. Row faces and intertwiners that lose support, and the alphabet $A_4$, are not covered.

\section{Beyond height alphabets}\label{sec:extensions}

The mechanism of Section~\ref{sec:support} applies to encodings that are not height alphabets in the sense of Definition~\ref{def:encoding}, namely to blocks of two sites and to the binary alphabet with $1\sim1$, on which the hard-core rule is carried by vanishing weights.

\subsection{Blocks of two sites}

Grouping two horizontally adjacent sites gives the three blocks $\bz=00$, $\bL=01$, $\bR=10$. For blocks $s=(s_1,s_2)$ and $t=(t_1,t_2)$ put
\[
 \mathsf H(s,t)=\mathbf 1\{s_2t_1=0\},\qquad
 \mathsf V(s,t)=\mathbf 1\{s_1t_1=s_2t_2=0\},
\]
the conditions for $t$ to be placed to the right of $s$, respectively below $s$. A \emph{blocked face} $\fc abcd$ is admissible if $\mathsf H(a,b)\mathsf H(c,d)\mathsf V(a,c)\mathsf V(b,d)=1$, that is, if the corresponding $2\times4$ binary pattern contains no two adjacent ones. Let $\cF$ be the set of admissible blocked faces, of which there are $41$ (Appendix~\ref{app:computations}), and let $n(\bz)=0$, $n(\bL)=n(\bR)=1$. The \emph{blocked hard-square face} with corner factors $A,B,C,D$ is
\begin{equation}\label{eq:blocked-hs-face}
 Q\fc abcd=A^{n(a)}B^{n(b)}C^{n(c)}D^{n(d)}\qquad\left(\fc abcd\in\cF\right).
\end{equation}
On a periodic lattice of blocks every block occupies each corner position exactly once, so the product of these weights is exactly the hard-square weight $z^{|x|}$ with $z=ABCD$, so blocking is an exact regrouping. Blocked face weights are functions on $\cF$, and \eqref{eq:FYB} is read with $g$ running over the three blocks.

\begin{theorem}[One-equation obstruction]\label{thm:blocked-singleton}
Let $Q$ be the blocked hard-square face with $B,C\neq0$, and let $P,R$ be blocked face weights satisfying \eqref{eq:FYB-rho} for some $\rho$. Then
\begin{equation}\label{eq:blocked-singleton}
 BC\;P\fc{\bL}{\bz}{\bz}{\bz}\,R\fc{\bL}{\bL}{\bz}{\bz}=0 .
\end{equation}
\end{theorem}

\begin{proof}
Take $(a,b,c,d,e,f)=(\bz,\bz,\bL,\bL,\bz,\bL)$. On the left of \eqref{eq:FYB-rho} the factors are $P\fc{\bL}{g}{\bz}{\bz}$, $Q\fc{\bz}{\bL}{\bL}{g}$ and $R\fc{\bL}{\bL}{g}{\bz}$. Admissibility of the $Q$-face requires $\mathsf H(\bL,g)=\mathsf V(\bL,g)=1$, that is $g_1=g_2=0$, so $g=\bz$; then all three faces are admissible and $Q\fc{\bz}{\bL}{\bL}{\bz}=BC$. On the right, the factor $P\fc{\bz}{\bL}{g}{\bL}$ is inadmissible for every $g$, because $\mathsf V(\bL,\bL)=0$.
\end{proof}

Since admissibility does not depend on the parity of a face, the same computation applies verbatim to parity-staggered blocked weights, with $P$ and $R$ replaced by the appropriate types.

\begin{corollary}[No analytic blocked Yang--Baxter family]\label{cor:blocked-analytic}
No analytic Yang--Baxter family of blocked face weights on $\cF$ passes through a blocked hard-square face with $A,B,C,D\neq0$. Its weights are assumed to vanish off $\cF$; no genericity assumption is made.
\end{corollary}

\begin{proof}
With $P=W_{p_0q}$ and $R=W_{qr_0}$, the product of the holomorphic functions $q\mapsto P\fc{\bL}{\bz}{\bz}{\bz}$ and $q\mapsto R\fc{\bL}{\bL}{\bz}{\bz}$ vanishes identically by Theorem~\ref{thm:blocked-singleton}. The first equals $A\neq0$ at $q=r_0$, and the second equals $AB\neq0$ at $q=p_0$, which is impossible on a connected domain.
\end{proof}

The proofs of Theorem~\ref{thm:blocked-singleton} and Corollary~\ref{cor:blocked-analytic} use only that $Q$ does not vanish on the faces $\fc{\bz}{\bL}{\bL}{\bz}$, $\fc{\bL}{\bz}{\bz}{\bz}$ and $\fc{\bL}{\bL}{\bz}{\bz}$. They therefore apply to every blocked face weight with full support on $\cF$, and in particular to every blocked face weight whose product over the faces of each periodic blocked configuration is $c^Nz^{|x|}$, with $N$ the number of faces and $c,z\neq0$, because, as in Proposition~\ref{prop:presentations}, every admissible blocked face occurs in such a configuration, whose weight is nonzero.

The blocked alphabet is not a height alphabet, since its horizontal and vertical rules differ and $\mathsf H$ is not symmetric. This affects the reading of \eqref{eq:FYB}. In the train argument the faces $P\fc fgab$ and $Q\fc edfg$ lie in consecutive rows, the top row of $P$ being the bottom row of $Q$, and $R$ is the intertwining face. Each of its edges is a column of $P$ or of $Q$ on one of the two sides of \eqref{eq:FYB}, so all of them are vertical edges of the lattice. For the blocked model the natural support of the intertwiner is therefore not $\cF$ but the set $\cF_{\mathsf V}$ of blocked faces whose four edges satisfy the vertical rule, $\mathsf V(a,b)\mathsf V(c,d)\mathsf V(a,c)\mathsf V(b,d)=1$. Theorem~\ref{thm:blocked-singleton} and Corollary~\ref{cor:blocked-analytic} concern instead weights on $\cF$ in all three positions, so that the edge $gb$, a column of $P$ and a row of $R$, carries both rules. They do not cover the transfer-matrix structure of the blocked model, with row faces on $\cF$ and intertwiners on $\cF_{\mathsf V}$. There $R\fc{\bL}{\bL}{\bz}{\bz}=0$, since $\mathsf V(\bL,\bL)=0$, so the equation of Theorem~\ref{thm:blocked-singleton} holds trivially, and the structure admits the solution $(Q,Q,\mathrm{Id})$, with $\mathrm{Id}\fc abcd=\delta_{b,c}$ on $\cF_{\mathsf V}$. A single family as in Definition~\ref{def:analytic-family} cannot have this structure, since $W_{qr_0}$ equals the blocked hard-square face at $q=p_0$, and that face does not vanish on $\fc{\bL}{\bL}{\bz}{\bz}\notin\cF_{\mathsf V}$; whether a family of row faces with intertwiners on $\cF_{\mathsf V}$ can pass through it is open (Question~\ref{q:blocks}). The argument still works if the weights, in all three positions, may also be nonzero on $\cF_{\mathsf V}$. In the boundary equation of Theorem~\ref{thm:blocked-singleton} the middle face still forces $g=\bz$ on the left, and on the right the face $P\fc{\bz}{\bL}{g}{\bL}$ has the column $(\bL,\bL)$, which violates the vertical rule, so it lies neither in $\cF$ nor in $\cF_{\mathsf V}$. Families that may be nonzero on still larger sets of blocked faces are not covered.

\subsection{Exclusion by vanishing weights}\label{sec:full-binary}

On the binary alphabet with $1\sim1$ every face is admissible, so the hard-core rule can be imposed only by the weights. Let $\widehat Q$ be the face weight equal to a pure hard-square face \eqref{eq:pure-hs-face} on the seven faces \eqref{eq:seven-labels} and to $0$ on the nine faces with two adjacent occupied sites. Its product over the faces of a torus is $z^{|x|}$ on hard-square configurations and $0$ on all others, so $\widehat Q$ reproduces hard squares although the alphabet does not satisfy Definition~\ref{def:encoding}(1). Auxiliary faces are now functions on all sixteen faces. We keep the labels $p_0,\dots,p_6$ and $r_0,\dots,r_6$ of \eqref{eq:seven-labels} and write
\[
 x=P\fc0101,\qquad v=P\fc1010,\qquad y=R\fc0011,\qquad u=R\fc1100 .
\]

\begin{proposition}[Binary alphabet with $1\sim1$]\label{prop:full-binary}
Let $P,R$ be face weights on the binary alphabet with $1\sim1$ that satisfy \eqref{eq:FYB-rho} for some $\rho\neq0$, with middle face $\widehat Q$ and $A,B,C,D\neq0$. Then $p_2p_4r_3r_4r_5=0$. Consequently no analytic Yang--Baxter family on this alphabet passes through $\widehat Q$.
\end{proposition}

\begin{proof}
For the boundary data of \eqref{eq:b1}--\eqref{eq:b6}, \eqref{eq:FYB-rho} gives the same equations as on the binary alphabet, with the factor $\rho$ on the right sides, and the boundary datum of \eqref{eq:b7} acquires the single term $-A\rho\,vu$, from the face $\fc1010$ of $P$ and the face $\fc1100$ of $R$. Five further boundary data give
\begin{align*}
 (0,0,0,0,0,0):&\quad Dp_3r_2+(1-\rho)p_0r_0-\rho Ap_2r_3=0,\\
 (1,0,1,1,0,1):&\quad BC(vu-\rho xy)=0,\\
 (1,0,1,1,0,0):&\quad B(p_2u-\rho Cxr_1)=0,\\
 (1,0,1,0,0,1):&\quad C(vr_3-\rho Bp_1y)=0,\\
 (1,0,1,0,0,0):&\quad p_2r_3+Dp_4r_4-\rho BCp_1r_1=0.
\end{align*}
As in the proof of Theorem~\ref{thm:binary-certificate}, the first six equations give $p_6r_5=Ap_1r_0$, $p_1r_6=Dp_0r_5$ and $p_6r_6=Dp_5r_5$, hence $Dr_5(p_5r_5-Ap_0r_0)=0$. With this, the seventh equation and $A$ times the eighth give $r_5(vu-Ap_2r_3)=0$, and the ninth turns this into $r_5(\rho xy-Ap_2r_3)=0$. Multiplying the tenth and eleventh equations, in the form $p_2u=\rho Cxr_1$ and $vr_3=\rho Bp_1y$, gives $p_2r_3\,vu=\rho^2BC\,p_1r_1\,xy$. By the ninth equation this is $\rho\,xy\,(p_2r_3-\rho BCp_1r_1)=0$, and by the twelfth it is $\rho D\,xy\,p_4r_4=0$. Multiplying by $r_5$ and using $\rho r_5xy=Ar_5p_2r_3$ gives $AD\,p_2p_4r_3r_4r_5=0$.

For the second claim, put $P=W_{p_0q}$ and $R=W_{qr_0}$. At $q=r_0$ the weights $p_2$ and $p_4$ equal $C$ and $BC$, and at $q=p_0$ the weights $r_3$, $r_4$ and $r_5$ equal $B$, $BC$ and $A$. Each of these five holomorphic functions of $q$ is therefore not identically zero on the connected domain $\mathcal D$, and neither is their product; but the product vanishes for every $q$.
\end{proof}

The proof uses only the values of $\widehat Q$ on the seven hard-square faces and its vanishing on the five faces $\fc1100$, $\fc0011$, $\fc0101$, $\fc0111$ and $\fc1110$. It does not cover face weights that kill the other configurations with fewer zeros, such as the pure hard-square face extended by zero only on the four faces with $a=c=1$ and on $\fc1100$, nor other alphabets in which the hard-core rule is imposed by vanishing weights. That face weight kills every configuration with two adjacent occupied sites. An occupied vertical pair fills the corners $a,c$ of a face, and an occupied horizontal pair fills $a,b$ of a face $\fc11cd$, which is one of these unless $(c,d)=(0,1)$, in which case the site below its right end gives a vertical pair.

\section{Hidden heights and open questions}\label{sec:conclusion}

Within height alphabets, the remaining possibility is an encoding with hidden heights, in which the number of height configurations over a hard-square configuration is unbounded, or depends on the configuration (for a connected alphabet over which every hard-square configuration is an image, a number that depends on the configuration is also unbounded, by the paragraph after Proposition~\ref{prop:classification}). The hard-square weights then appear only after summing over the hidden heights, as in Proposition~\ref{prop:fibre-sums}, and Propositions~\ref{prop:presentations} and~\ref{prop:fibre-sums}, which identify the face on which the certificates act, are not available.

\begin{question}\label{q:hidden}
Is there an analytic Yang--Baxter family of face weights on a height alphabet with hidden heights whose shadow, at some value of the spectral parameters, is the hard-square model with nonzero activity? A natural first test is the restricted solid-on-solid alphabets $A_{r-1}$, $r\ge6$, with occupation maps compatible with the hard-core rule, and the alphabets of the dilute $A_L$ models \cite{WarnaarNienhuisSeaton92}.
\end{question}

\begin{question}\label{q:blocks}
Are there analytic families of blocked faces $W_{uv}$ on $\cF$ and $S_{uv}$ on $\cF_{\mathsf V}$ such that every triple $(W_{pq},W_{pr},S_{qr})$ satisfies \eqref{eq:FYB-rho} and some $W_{p_0r_0}$ is the blocked hard-square face? This is the transfer-matrix structure of the blocked model, which Corollary~\ref{cor:blocked-analytic} does not cover (Section~\ref{sec:extensions}). Does Corollary~\ref{cor:blocked-analytic} extend to blocks of $k\ge3$ sites, and to blocks of arbitrary shape?
\end{question}

\appendix
\section{Finite computations}\label{app:computations}

The finite statements listed below are reproduced by the ancillary script \nolinkurl{anc/verify_certificates.py}, which is also available at \url{https://github.com/gabrielortegaproyectos/hard-square-yang-baxter-obstruction} and archived as version~1.0.0 at \href{https://doi.org/10.5281/zenodo.23892182}{\nolinkurl{doi:10.5281/zenodo.23892182}}. It is written in Python~3 with the libraries \texttt{sympy} and \texttt{networkx}, and tested with Python~3.12, \texttt{sympy}~1.14.0 and \texttt{networkx}~3.7. Checks are exact (rational arithmetic in \texttt{sympy}, or integer combinatorics) unless they are marked as floating point. The default run takes under a minute; the options \texttt{--groebner} and \texttt{--classify} add two longer cross-checks. Every proof in the paper can be read without the script, which only guards against arithmetic slips.

\subsection*{Binary alphabet (Section~\ref{sec:support})}
\begin{enumerate}
\item Enumeration of the $64$ boundary data of \eqref{eq:FYB} with middle face \eqref{eq:pure-hs-face}. Exactly $18$ equations are not identically zero, and the equations quoted in Section~\ref{sec:support} agree with the enumeration.
\item The combinations producing $E_1,\dots,E_4$, and the identity \eqref{eq:binary-certificate}. With $E_5=Dp_3r_2-Ap_2r_3$, the equation for $(0,0,0,0,0,0)$, the identity $A^2D\,r_5p_2r_3=r_6E_1+Ar_0E_2-r_5E_3-Dr_5E_4-ADr_5E_5$, which is \eqref{eq:a4-certificate} for unstaggered weights and $\rho=1$ (remark after Theorem~\ref{thm:a4-certificate}).
\item The identities of Remark~\ref{rem:orientations} for an equation with a scalar factor.
\item Substitution of \eqref{eq:branch-r} and \eqref{eq:branch-p} into all $18$ equations, and the uniqueness of the solution on each of the two supports, which is \eqref{eq:branch-r}, respectively \eqref{eq:branch-p}; up to corner factors with product $1$, the two branches are $(Q,\mathrm{Id})$ and $(\mathrm{Id},Q)$ (remark after Theorem~\ref{thm:near-full}).
\item With \texttt{--groebner}, for every set $Z$ of at most two auxiliary weights, the ideal generated by the equations with the weights in $Z$ set to zero, saturated by the product of the remaining weights, is the unit ideal over $\Q(A,B,C,D)$ unless $Z=\{p_2,p_3\}$ or $Z=\{r_2,r_3\}$. The proof of Theorem~\ref{thm:near-full} does not depend on this.
\end{enumerate}

\subsection*{The alphabet $A_4$ (Sections~\ref{sec:prelim} and~\ref{sec:support})}
\begin{enumerate}
\item (Floating point.) The weights of Example~\ref{ex:rsos-calibration} satisfy \eqref{eq:FYB}, with $P=W_u$, $Q=W_{u+v}$ and $R=W_v$, for four pairs $(u,v)$, with residuals below $10^{-15}$; transported to occupations by \eqref{eq:types}, they satisfy \eqref{eq:staggered}. This calibrates the convention of Definition~\ref{def:fyb} against a known integrable model.
\item Admissible $A_4$ faces of each type correspond bijectively to the seven admissible binary faces.
\item Lemma~\ref{lem:staggered}, computed from the heights. For $P,Q,R$ given type by type by \eqref{eq:types} with symbolic weights, the equation \eqref{eq:FYB-rho} for each of the $4^6$ boundary data of $A_4$ vanishes unless the six sides of the hexagon join adjacent heights, and is otherwise the equation \eqref{eq:staggered} indexed by the parity of $a$ and the occupations of the six heights. With middle face $(s_0Q^0,s_1Q^1)$, exactly $36$ of these equations are not identically zero, $18$ of each parity $\alpha$, and each is $s_\alpha$ times the equation \eqref{eq:staggered} with $s_0=s_1=1$ and $\rho_\alpha=\rho\,s_{1-\alpha}/s_\alpha$ (Section~\ref{sec:support}).
\item With a staggered pure hard-square middle face and type-dependent factors $\rho_0,\rho_1$, exactly $36$ of the $128$ equations \eqref{eq:staggered} are not identically zero, and \eqref{eq:s1}--\eqref{eq:s8} agree with the enumeration.
\item The combinations producing $G_1,\dots,G_5$, the identity \eqref{eq:a4-certificate}, and the invariance of the system under the exchange of the two types. The eight symmetries of the square map staggered pure hard-square faces to staggered pure hard-square faces, exchanging the types exactly when they move the top-left corner to $b$ or $c$ (remark after Theorem~\ref{thm:a4-certificate}).
\item On the tori $T_2$ and $T_4$ there are $7$ and $743$ hard-square configurations and $14$ and $1486$ height configurations on $A_4$, and every fibre of the occupation map is an orbit of $\sigma$ of size two (Example~\ref{ex:encodings}(2)).
\item On $T_6$ and $T_8$, the weights of the lifts $h_x$ and $\sigma\circ h_x$ of the configurations used in the proof of Proposition~\ref{prop:fibre-sums} (a single particle, two distant particles, a diagonal pair of each kind, and two distant diagonal pairs of each kind) are the stated monomials in the weights of $W^0$ and $W^1$.
\item On cylinders of circumference $M=4,6$, the identity $T_1T_0G_1=G_1(C_MZ_M)^2$ holds for a staggered pure hard-square face with $s_0=s_1=1$ and symbolic corner factors (Remark~\ref{rem:why-class}).
\item With \texttt{--classify}, consider the $86\,828$ connected height alphabets on the labelled heights $\{0,\dots,N-1\}$ with $2\le N\le5$, with at least one occupied and one empty height, and with no two occupied heights adjacent (so loops occur only at empty heights). Alphabets without occupied heights or without empty heights, among them those with a single height, have no height configuration over a single particle or over the empty configuration, and are left out. Among these alphabets, those for which the empty configuration, a single particle, a full sublattice, the four neighbours of a site and three two-particle configurations on the $4\times4$ torus all have the same number of preimages, at most $20$, form exactly two isomorphism classes; compared by isomorphism of graphs with occupations, they are the binary alphabet and $A_4$. The bound $20$ plays the role of $\sup_M\kappa_M$ in Definition~\ref{def:encoding}(2), and the enumeration of preimages stops once it is exceeded. This illustrates Proposition~\ref{prop:classification}, which is proved in the text.
\item Transported to occupations, the weights of Example~\ref{ex:rsos-calibration} are the same for both types (compared exactly at $u=0$ and $u=\lambda$), and at $u=-\lambda$ they satisfy $w_6=0$, $w_4w_0=w_3w_2$ and $z=(11+5\sqrt5)/2$.
\end{enumerate}

\subsection*{Blocks and the binary alphabet with $1\sim1$ (Section~\ref{sec:extensions})}
\begin{enumerate}
\item The $41$ admissible blocked faces, and the equivalence of blocked admissibility with admissibility of the corresponding $2\times4$ binary pattern.
\item With middle face \eqref{eq:blocked-hs-face}, the equation for $(\bz,\bz,\bL,\bL,\bz,\bL)$ is \eqref{eq:blocked-singleton}, also when $P,R$ may be nonzero on $\cF\cup\cF_{\mathsf V}$; the set $\cF_{\mathsf V}$ has $35$ faces, and $\cF\cup\cF_{\mathsf V}$ has $49$.
\item On the binary alphabet with $1\sim1$ and middle face $\widehat Q$, the twelve equations of the proof of Proposition~\ref{prop:full-binary} agree with the enumeration, and $\rho A^2BCD\,p_2p_4r_3r_4r_5$ is a polynomial combination of them with the multipliers of the proof. The twelve equations do not change when $\widehat Q$ takes arbitrary values on $\fc1010$, $\fc1011$, $\fc1101$ and $\fc1111$, and each of the five zeros of the remark after Proposition~\ref{prop:full-binary} enters one of them. The face weight extended by zero only on the four faces with $a=c=1$ and on $\fc1100$ vanishes on some face of a configuration on $T_2$, $T_3$ or $T_4$ exactly when two adjacent sites are occupied.
\end{enumerate}

\section*{Acknowledgements}
The computer-algebra verifications described in Appendix~\ref{app:computations} are reproducible with the ancillary script \nolinkurl{anc/verify_certificates.py}, archived at \href{https://doi.org/10.5281/zenodo.23892182}{\nolinkurl{doi:10.5281/zenodo.23892182}}. Generative AI tools were used to assist in drafting, in cross-checking proofs, and in writing the verification scripts; the authors take full responsibility for the content.

\end{document}